\documentclass[aps,prd,superscriptaddress,raggedfooter,raggedbottom,amssymb,amsmath]{revtex4-2}
\usepackage{graphicx}
\usepackage{subfigure}
\usepackage{color}
\usepackage{mathrsfs}
\usepackage{float}
\usepackage{soul}

\usepackage[breaklinks=true,colorlinks=true]{hyperref}

\hypersetup{colorlinks=true,citecolor=blue,linkcolor=blue,urlcolor=blue}

\DeclareMathOperator\sgn{sgn}

\begin{document}

\title{Kink-Antikink Collisions and Multi-Bounce Resonance Windows in Higher-Order Field Theories}

\author{Ivan C.\ Christov}
\affiliation{School of Mechanical Engineering, Purdue University, West Lafayette, Indiana 47907, USA}

\author{Robert J.\ Decker}
\affiliation{Mathematics Department, University of Hartford, 200 Bloomfield Ave., West Hartford, CT 06117, USA}

\author{A.\ Demirkaya}
\affiliation{Mathematics Department, University of Hartford, 200 Bloomfield Ave., West Hartford, CT 06117, USA}

\author{Vakhid~A.~Gani}
\affiliation{Department of Mathematics, National Research Nuclear University MEPhI (Moscow Engineering Physics Institute), Moscow 115409, Russia} 
\affiliation{Theory Department, Institute for Theoretical and Experimental Physics of National Research Centre ``Kurchatov Institute'', Moscow 117218, Russia
}

\author{P.~G.\ Kevrekidis}
\affiliation{Department of Mathematics and Statistics, University of Massachusetts, Amherst, MA 01003-4515, USA}
\affiliation{Mathematical Institute, University of Oxford, Oxford OX26GG, UK}

\author{Avadh Saxena}
\affiliation{Theoretical Division and Center for Nonlinear Studies, Los Alamos National Laboratory, Los Alamos, New Mexico 87545, USA}

\begin{abstract}
We study collisions of coherent structures in higher-order field-theoretic models, such as the $\phi^8$, $\phi^{10}$ and $\phi^{12}$ ones. The main distinguishing feature, of the example models considered herein, is that the collision arises due to the long-range interacting algebraic tails of these solitary waves. We extend the approach to suitably initialize the relevant kinks, in the additional presence of finite initial velocity, in order to minimize the dispersive wave radiation potentially created by their slow spatial decay. We find that, when suitably initialized, these models still feature the multi-bounce resonance windows earlier found in models in which the kinks bear exponential tails, such as the $\phi^4$ and $\phi^6$ field theories among others. Also present is the self-similar structure of the associated windows with three- and more-bounce windows at the edges of two- and lower-bounce ones. Moreover, phenomenological but highly accurate (and predictive), scaling relations are derived for the dependence of the time between consecutive collisions and, e.g., the difference in kinetic energy between the incoming one and the critical one for one-bounces. Such scalings are traced extensively over two-bounce collision windows throughout the three models, hinting at the possibility of an analytical theory in this direction.
\end{abstract}

\maketitle

\section{Introduction}
\label{sec:Introduction}

Topological solitons play an important role in modern physics \cite{Vilenkin.book.2000,Manton.book.2004,Vachaspati.book.2006}. In this regard, for many decades, field-theoretic models in $(1+1)$-dimensional space-time have been actively studied. In addition to being of particular  interest in their own right, such models are relevant to a wide range of physical applications~\cite{sg,Kevrekidis.book.2019,Bishop.PhysD.1980}. The understanding of models with one real scalar field with both polynomial and non-polynomial  self-interaction (potential) is continuing to grow. More specifically, methods of constructing new models with interesting properties, such as the deformation procedure \cite{Bazeia.PRD.2002,Bazeia.PRD.2004,Bazeia.AP.2018,Blinov.arXiv.2020.deform,Blinov.JPCS.2020.deform} have been proposed, parametric field theories have been considered~\cite{Demirkaya.JHEP.2017}, and ideas such as those involving quasi-normal modes have been used to explain dynamical model features~\cite{Dorey.PLB.2018}.  Additionally, topologically nontrivial solutions are actively investigated in models with two real scalar fields \cite{Gani.YaF.2001.SuSy,Alonso-Izquierdo.AHEP.2013,Katsura.PRD.2014,Alonso-Izquierdo.PS.2019,Alonso-Izquierdo.CNSNS.2019}, as well as in models with a richer set of fields \cite{Gani.JHEP.2016,Klimashonok.PRD.2019,Perapechka.PRD.2020}. As some of the prototypical uses of these field theories one can mention toy models for dark matter halos~\cite{prd12}, as well as cosmological applications of the Higgs field~\cite{prd13}.

Among the topological defects in $(1+1)$-dimensional field-theoretic models, a special place is reserved for {\it kinks} --- topologically nontrivial field configurations connecting degenerate minima of the model's potential \cite[Chap.~5]{Manton.book.2004}. Kinks arise in various physical processes and represent coherent structures in the form of heteroclinic connections. For example, in cosmology a flat domain wall separating regions of space with different vacua, in the direction perpendicular to the wall, is a kink-type field configuration \cite{Vachaspati.book.2006,Ahlqvist.2015}. In addition, kinks arise in models of defects within crystals and in various phenomena in graphene \cite{Killi.IJMPB.2012,Yamaletdinov.PRB.2017,Yamaletdinov.Carbon.2019,Yamaletdinov.Chapter.2021}. The effort to understand the processes of kink interactions, such as collisions and scattering thus has a time-honored history, detailed, e.g., recently in~\cite{Kevrekidis.book.2019}.

Starting with the work of Ref.~\cite{Kudryavtsev.JETPLett.1975} it was noted that, in the collision of the kink and antikink of the $\phi^4$ model at small initial velocities (therein $v_{\rm in}^{}=0.1$ is in units of the speed of light), the kinks are captured and form a ``bion,'' which is a bound state of a kink and an antikink. Subsequently, the bion slowly decays. This situation was somewhat at odds with the naive idea that the kink and antikink should annihilate upon colliding, immediately disintegrating into radiation of small-amplitude waves, which take away the energy of the colliding kinks. It is also fundamentally distinct from the elastic kink-antikink collision featured in integrable models such as the sine-Gordon equation~\cite[Sec.~5.3]{Manton.book.2004}. Nevertheless, it is a natural manifestation of the non-integrability of this partial differential equation (PDE). Then, it was found that there is a critical value $v_{\rm cr}^{}$ of the initial velocity ($v_{\rm cr}^{}\approx 0.26$ for kinks of the $\phi^4$ model) separating two different regimes of the collision: for $v_{\rm in}^{}<v_{\rm cr}^{}$, the capture and formation of a bound state occurs, while for $v_{\rm in}^{}>v_{\rm cr}^{}$ the kink and antikink escape to infinity after their collision, see, e.g., Ref.~\cite{Sugiyama.PTP.1979}.

Further investigation of this phenomenon led to the discovery of {\it escape windows}. It was found that in the range $v_{\rm in}^{}<v_{\rm cr}^{}$ there are intervals of the initial velocities within which kinks scatter and eventually escape to the spatial infinities. The difference from the case of $v_{\rm in}^{}>v_{\rm cr}^{}$, however, is that inside the escape windows the kinks scatter to infinity not after a single impact, but after two or more successive collisions. This phenomenon has been explained on the basis of a resonant energy exchange mechanism \cite{Campbell.PhysD.1983,Campbell.PhysD.1986,Goodman.JADS.2005,Goodman.Chaos.2015}, see also Ref.~\cite{Belova.UFN.1997} for a review. The essence of the mechanism is as follows. In the first collision, part of the kinetic energy of the kinks is transferred into the excitation of the vibrational mode of each kink. After that, the kinks scatter, but because of the decreased kinetic energy they are unable to overcome mutual attraction and go to infinity. Instead, they stop at some distance and then return to collide again. In the second collision, provided some resonance condition is fulfilled, part of the energy of the vibrational mode can be returned to the kinetic energy of the kinks, which allows them to escape to infinity. The intervals of the initial velocity from the range $v_{\rm in}^{}<v_{\rm cr}^{}$, within which the kinks escape to infinity after two collisions, were named {\it two-bounce escape windows}. Note that the resonant energy return from the vibrational mode of kinks to their kinetic energy can occur not only in their second collision, but in the third, fourth, and further collisions. Intervals of the initial velocities, within which kinks escape to infinity after three, four, and further collisions were named {\it three-bounce, four-bounce}, and so on escape windows. The resonance condition mentioned above relates the frequency of the vibrational mode of the kink and the time interval between the first and second collisions of kinks (for two-bounce escape windows), or between the second and third collisions of kinks (for three-bounce escape windows), and so on. It turned out that the frequency of the vibrational mode of the $\phi^4$ kink fits very well into this picture.

Similar scenarios were also found in the kink-antikink collisions in other field-theoretic models, e.g., in the $\phi^6$ model \cite{Dorey.PRL.2011,Gani.PRD.2014}, the modified sine-Gordon \cite{Peyrard.PhysD.1983.msG}, the double sine-Gordon \cite{Campbell.PhysD.1986.dsG,Gani.PRE.1999,Gani.EPJC.2018}, and the sinh-deformed $\phi^4$ model \cite{Bazeia.EPJC.2018}. However, noticeable deviations from this clear and rather simple phenomenology, developed on the basis of the $\phi^4$ kinks' scattering, have emerged. In the scattering of kinks of the double sine-Gordon model, the so-called quasiresonances were found \cite[Fig.~20]{Campbell.PhysD.1986.dsG} in suitable parametric regimes, and maxima in the dependence of the time $T_{23}$ between the second and third collisions of kinks on the initial velocity  were observed instead of some two-bounce escape windows. It was shown that quasiresonances and escape windows are phenomena of the same nature, and for some values of the model parameter they coexist \cite[Fig.~3]{Gani.PRE.1999}. In the same model, for some values of the model parameter, a deviation of the resonance frequency from the frequency of the vibrational mode of a solitary kink was also found. In Ref.~\cite{Gani.PRE.1999}, a mechanism was proposed that qualitatively and semi-quantitatively explains the frequency shift of the vibrational mode, based on the mutual influence of the vibrational modes of the kink and antikink when the modes are localized near the continuum.

It is important to emphasize that, despite its time-honored history, the interest in the processes of kink collisions
remains undiminished. In recent years, many new and interesting results have been obtained. In particular, in collisions of kinks of the double sine-Gordon model, a nontrivial dependence of the critical velocity on the model parameter was found \cite{Gani.EPJC.2018}, which turned out to be more complicated than it was originally obtained in \cite{Campbell.PhysD.1986.dsG}. The excitation spectrum of the kink of the particular $\phi^6$ model explored in Refs.~\cite{Dorey.PRL.2011,Gani.PRD.2014} does not have a vibrational mode at all. Nevertheless, in antikink-kink collisions, escape windows appeared. An explanation of this phenomenon was proposed in Ref.~\cite{Dorey.PRL.2011} based on the fact that the kinks of this $\phi^6$ model are asymmetric. A final state that can be classified as a bound state of {\it oscillons} --- localized oscillating structures behaving like particles --- was observed in the collisions of kinks of the double sine-Gordon model \cite{Gani.EPJC.2018}, as well as of the sinh-deformed $\phi^4$ model \cite{Bazeia.EPJC.2018} and hyperbolic models \cite{Bazeia.PLB.2020}. Both the bound states of two oscillons and the escape of oscillons to spatial infinity have been observed.

In Ref.~\cite{Alonso-Izquierdo.arXiv.2020.wobbling} the scattering of the wobbling $\phi^4$ kink and antikink has been investigated numerically. Some interesting changes in the fractal structure of escape windows were found. We also mention Ref.~\cite{Mohammadi.CNSNS.2020}, which deals with the scattering process of the kink and antikink of the model, which is a periodic modification of the $\phi^4$ model. In addition, a recent paper \cite{Bazeia.IJMPA.2019} studies the kink-antikink scattering in the hyperbolic $\phi^4$ and $\phi^6$ models. We also note several works in which various modifications of the $\phi^4$ model are considered in connection with kink interactions \cite{Dorey.PLB.2018,Adam.PRD.2020,Adam.arXiv.2019.weakly,Zhong.JHEP.2020,Yan.PLB.2020,bazeia}. 
Even the fundamental analysis of corresponding collective coordinate descriptions is still an active topic of  consideration~\cite{weig,clisthenis}, while reviews offer a summarizing view of this ever-growing field~\cite{lizunova}. The processes of collisions of more than two kinks, also called multikink scattering, are also under investigation \cite{Saadatmand.PRD.2015,Moradi.JHEP.2017,Moradi.EPJB.2017,Gani.EPJC.2019}.

The subject of the present work is kinks with power-law asymptotics (or, as they are sometimes referred to, power-law tails) \cite{our_PRD,our_PRL,Belendryasova.CNSNS.2019,Khare.JPA.2019,Manton.JPA.2019,Gomes.PRD.2012,Bazeia.JPC.2018,Mello.PLA.1998} and, in particular, their scattering properties upon collision with each other starting with finite (\emph{nonzero}) initial velocity. In all the field-theoretic models mentioned above, the kink solutions have exponential asymptotics approaching the spatial infinity. This means that the field approaches the vacuum value exponentially with increasing distance from the center of the kink. However, there are models (with both polynomial \cite{Khare.PRE.2014} and non-polynomial potentials \cite{Bazeia.JPC.2018}) in which kinks exhibit one- or two-sided power-law asymptotics. This situation leads to new physics compared to the exponential-tails case. For kinks with power-law tails the energy density of the kink is only weakly localized, i.e., a large part of the kink's mass is concentrated in its power-law tail, compared to the exponential-tail case. Importantly, long-range interaction between kinks occurs when their power-law tails overlap. This, in turn, implies that, in numerical experiments, the kinks are practically always interacting, and one needs to be especially careful as to how to initialize the relevant configurations. In other words, the ansatz forms that are usually employed as initial conditions for the numerical simulation of kink-antikink collisions are not applicable to problems involving kinks with power-law tails.

In Ref.~\cite{our_PRD}, it was shown that the use of the initial configuration (at $t=0$) in the form of the sum of the kink and antikink centered at $x=\pm x_0^{}$ and ``facing'' each other by their power-law tails, leads to significant disturbances arising for $t>0$ and distorting the picture of the kink-antikink interaction. In particular, the illusion of repulsion between a kink and an antikink may arise \cite{Belendryasova.CNSNS.2019}, while they are attracted to each other, in fact. In Ref.~\cite{our_PRD}, employing examples of $\phi^8$, $\phi^{10}$ and $\phi^{12}$ polynomial field-theoretic models, we proposed a {\it minimization procedure} that allows one to obtain an effectively static (in that the kink and antikink initially bear vanishing speed) configuration of the type of ``kink+antikink'' suitable for numerical simulations. This configuration enabled the study of the interaction of an initially static kink and its corresponding antikink, without the generation of numerical artifacts caused by a naive ansatz selection. However, the prior work has not addressed the {\it collisional dynamics} of kinks interacting via power-law tails. Therefore, within the context of higher-order field theories, the goal of the present study is a systematic investigation of collisions of a kink and an antikink, exhibiting long-range interactions via power-law tails, and starting with nonzero initial velocities. Specifically, we seek to address the multi-bounce resonance windows in these higher-order models and provide an interpretation of the associated phenomenology. Note that the $\phi^8$ field theory is a model for describing massless mesons with long-range interactions~\cite{lohe}.  Similarly, $\phi^{10}$ and $\phi^{12}$ field theories are necessary to capture multiple successive phase transitions, within the Ginzburg--Landau phenomenological theory of the latter, as discussed in Ref.~\cite{Khare.PRE.2014} and those therein.

It is relevant to note in passing that the presence of escape windows suggests the possibility of a resonant energy exchange between the kink translational mode and some vibrational mode. 
Indeed, in the realm of the classical $\phi^4$ model, 
a large fraction of studies on multi-bounce windows have
focused on the study of simplified ordinary differential equation (ODE) models that could
capture the main features of the relevant 
phenomenology~\cite{Sugiyama.PTP.1979,Campbell.PhysD.1983,Anninos.PRD.1991,Goodman.JADS.2005,Goodman.Chaos.2015}.
Nevertheless, recent works~\cite{Weigel.JPCS.2014,Takyi.PRD.2016} have
raised relevant concerns about the self-consistency of
such reductions and, indeed, after about 4 decades
of efforts, the fundamental question of the derivation
of such reduced models and of their ability to
quantitatively capture multi-bounce collisions
remains, effectively, open, to the best of our
understanding. This is also confirmed by the ongoing
recent efforts in this direction~\cite{weig,clisthenis}.
On the other hand, a kink with one or two power-law tails does {\it not} have a vibrational mode in the discrete spectrum, see, e.g., \cite[Sec.~4]{Belendryasova.CNSNS.2019} or \cite[Sec.~3]{Bazeia.JPC.2018}. In particular, the continuous spectrum exists throughout the imaginary axis of the spectral plane and the only localized eigenmode is associated with translational invariance (with an associated eigenfunction corresponding to the spatial derivative of the solution, i.e., the Goldstone mode). In view of these considerations,
we will not pursue such a collective coordinate description
herein.

This paper is organized as follows. In Section \ref{sec:Overview}, we present the example model potentials of the $\phi^8$, $\phi^{10}$, and $\phi^{12}$ type having kinks with power-law tails, which will be studied here, and also describe the method used to perform the numerical simulations of kink-antikink collisions. In Section \ref{sec:Initial_Conditions}, we discuss the construction of initial conditions for numerical simulation of the kink-antikink collisions starting from nonzero initial velocities. Then, in Section \ref{sec:V_in-V_out}, we summarize the results of a detailed study of the kink-antikink collisions in the chosen $\phi^8$, $\phi^{10}$, and $\phi^{12}$ models, taking into account the effect of nonzero initial kink velocity. We also compare the results obtained by using different initial conditions. In Section \ref{sec:Correspondence}, we  establish a one-to-one correspondence between the initial velocities of the colliding kinks at various initial separation distances, including an infinitely large one. Finally, Section \ref{sec:Conclusion} provides a summary of our findings and a number of directions for future work.

\section{Overview of the Mathematical Model and the Numerical Approach}
\label{sec:Overview}

\begin{figure}[t!]
\centering
\includegraphics[width=0.46\textwidth]{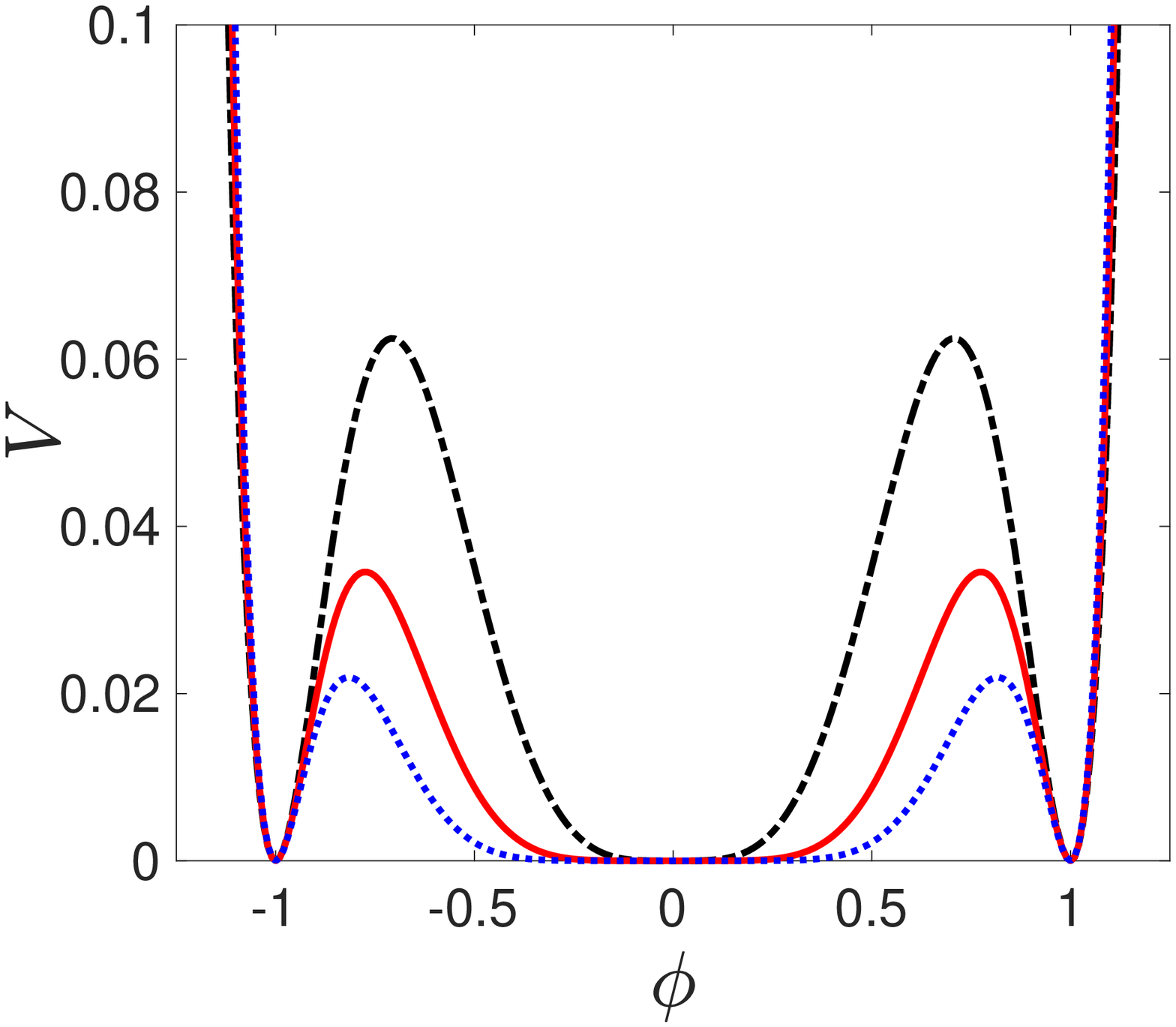}
\includegraphics[width=0.46\textwidth]{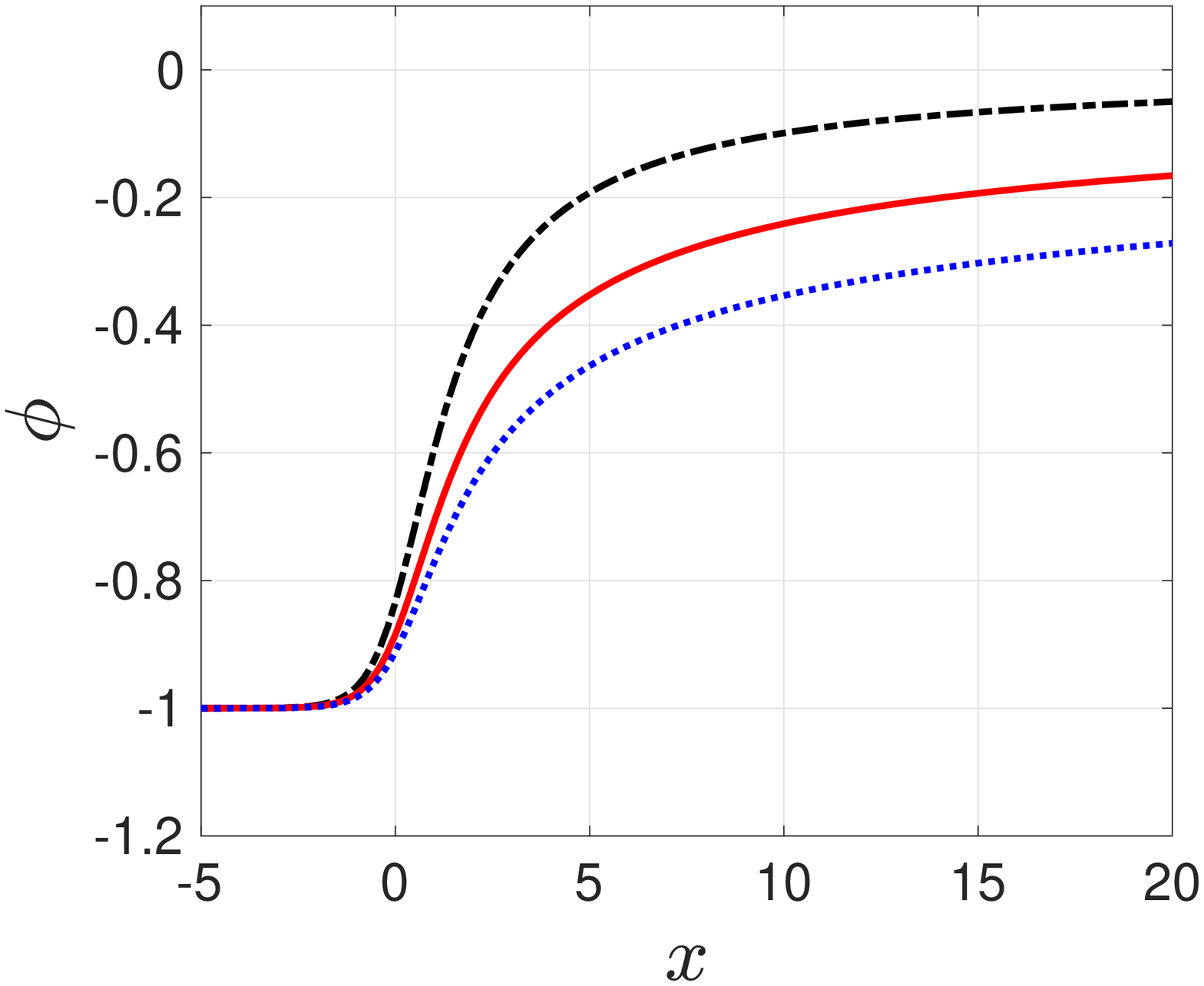}
\caption{The left panel presents the potential functions given in Eq.~\eqref{eq:V} and the right panel shows the kink solutions centered at $x=0$. In both panels, we have $m=2$ (black/dash-dotted curve), $m=3$ (red/solid curve), $m=4$ (blue/dotted curve).}
\label{Potential}
\end{figure}

Our aim is to simulate collisions between kinks and antikinks (``K-AK'' collisions) in $\phi^{2m+4}$ models for $m=2,3,4$, for which the potential has the functional form
\begin{equation}\label{eq:V}
    V(\phi) = \phi^{2m}(1-\phi^2)^2 \quad \mbox{for} \quad m=2,3,4.
\end{equation}
In keeping with the traditional notation, these models will be referred to as $\phi^8$, $\phi^{10}$, and $\phi^{12}$ based on the highest-order term in the polynomial potential, respectively, as shown in Fig.~\ref{Potential}(a). All kink solutions (kinks and antikinks) of these models have one power-law and one exponential tail, see \cite[Sec.~2]{our_PRD} for more details. For the models considered in this work, without loss of generality, the kinks and antikinks connect the constant solutions (vacua) $\phi=-1$ and $\phi=0$, see Fig.~\ref{Potential}(b).

Denote by $u=u(x,t)$ a real scalar field. Then, its dynamics, given a potential of the form in Eq.~\eqref{eq:V}, is described by an equation of motion obtained from Hamilton's principle, namely the PDE
\begin{equation}\label{eq:EoM}
    \frac{\partial^2 u}{\partial t^2} = \frac{\partial^2 u}{\partial x^2} - V^\prime(u).
\end{equation}
For our numerical simulations based on Eq.~\eqref{eq:EoM} we use a Fourier-based spectral method \cite{trefethen} with $N=500$ nodes in the $x$-direction in the interval $x\in[-L,+L]$ with $L=50$, making the $x$-increment $\Delta x=\frac{2(50)}{500}=0.2$. Discretization in $x$ yields a semi-discrete system of ODEs, which are integrated in $t$ using \textsc{Matlab}'s {\tt ode45} subroutine, which employs an adaptive-step-size method with specified tolerances (rather than a step size). The tolerances used for \textsc{Matlab}'s {\tt ode45} subroutine are $10^{-12}$ for both the absolute and relative tolerances.  We have added damping at the edges of the $x$ interval, using an infinitely differentiable \textquotedblleft bump\textquotedblright\ function (damping starts $20$ units from the boundaries of the interval, that is, from $x=\pm 30$ to $x=\pm 50$). The damping function has compact support; it is exactly zero in the interval $x\in[-30,30]$, and then rises to a value of five on the intervals $x\in[-50,-30]$ and $x\in[30,50]$ (thus there is no effect on any motion for $|x|\leq 30$). There is no ``corner'' at $x=\pm 30$ due to the damping function being infinitely differentiable with zero derivative there; this minimizes any radiation bounce. We track the position of a kink, which we label $x_{\rm c}^{}(t)$, as in \cite{our_PRD}. Briefly, we define $x_{\rm c}^{}(t)$ as the $x$ position such that $u\big(x_{\rm c}(t),t\big) =\tilde{\phi}$, where $\tilde{\phi}$ is $-0.83356$ for the $\phi^8$ model, $-0.884414$ for the $\phi^{10}$ model, and $-0.911608$ for the $\phi^{12}$ model. These values were chosen because they represent $\phi(x=0)$ for a single kink when using the implicit formulae from \cite{Khare.PRE.2014}.

\begin{figure}[t!]
\centering
\subfigure[]{\includegraphics[width=0.46\textwidth]{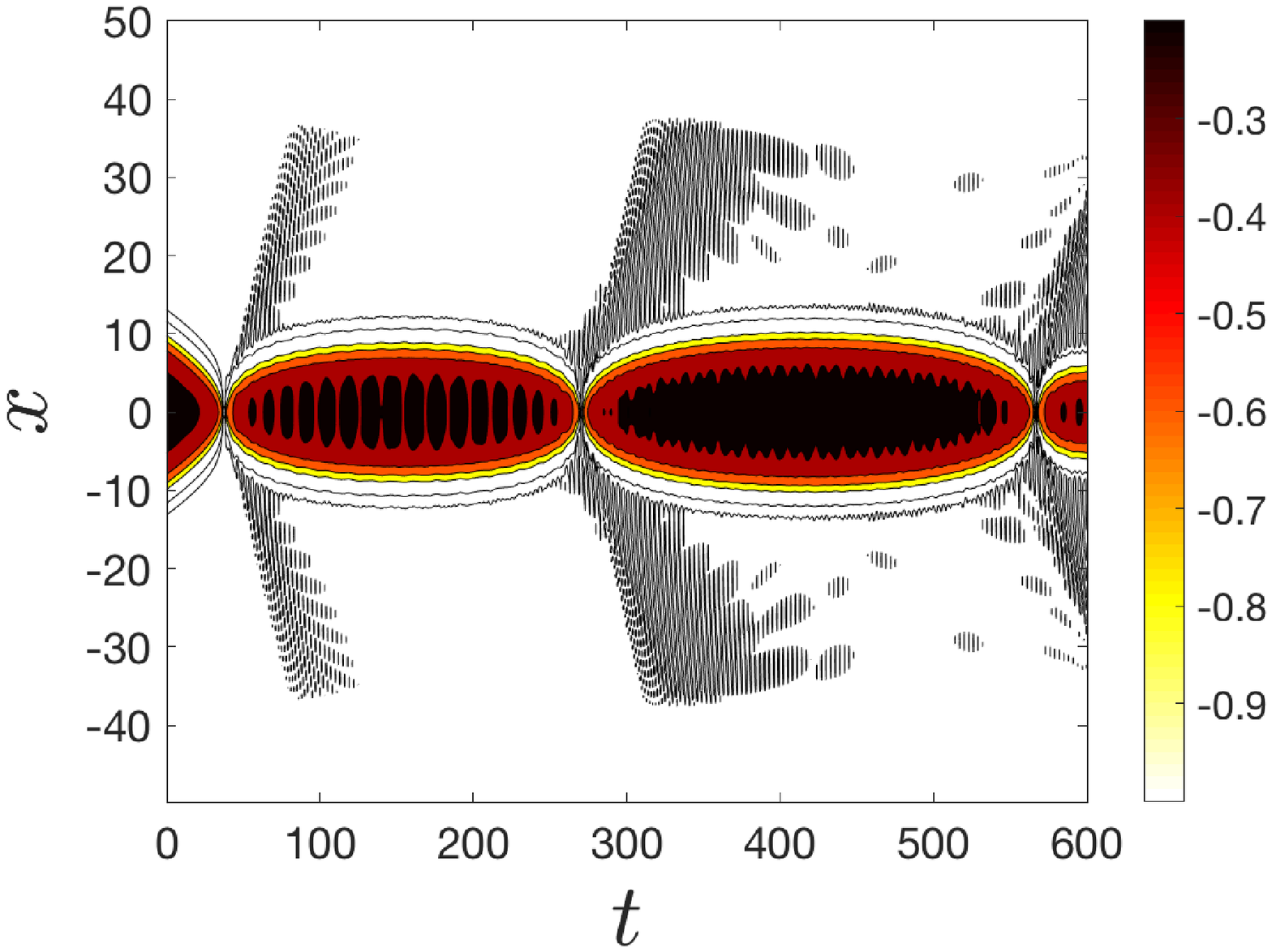}}%
\subfigure[]{\includegraphics[width=0.46\textwidth]{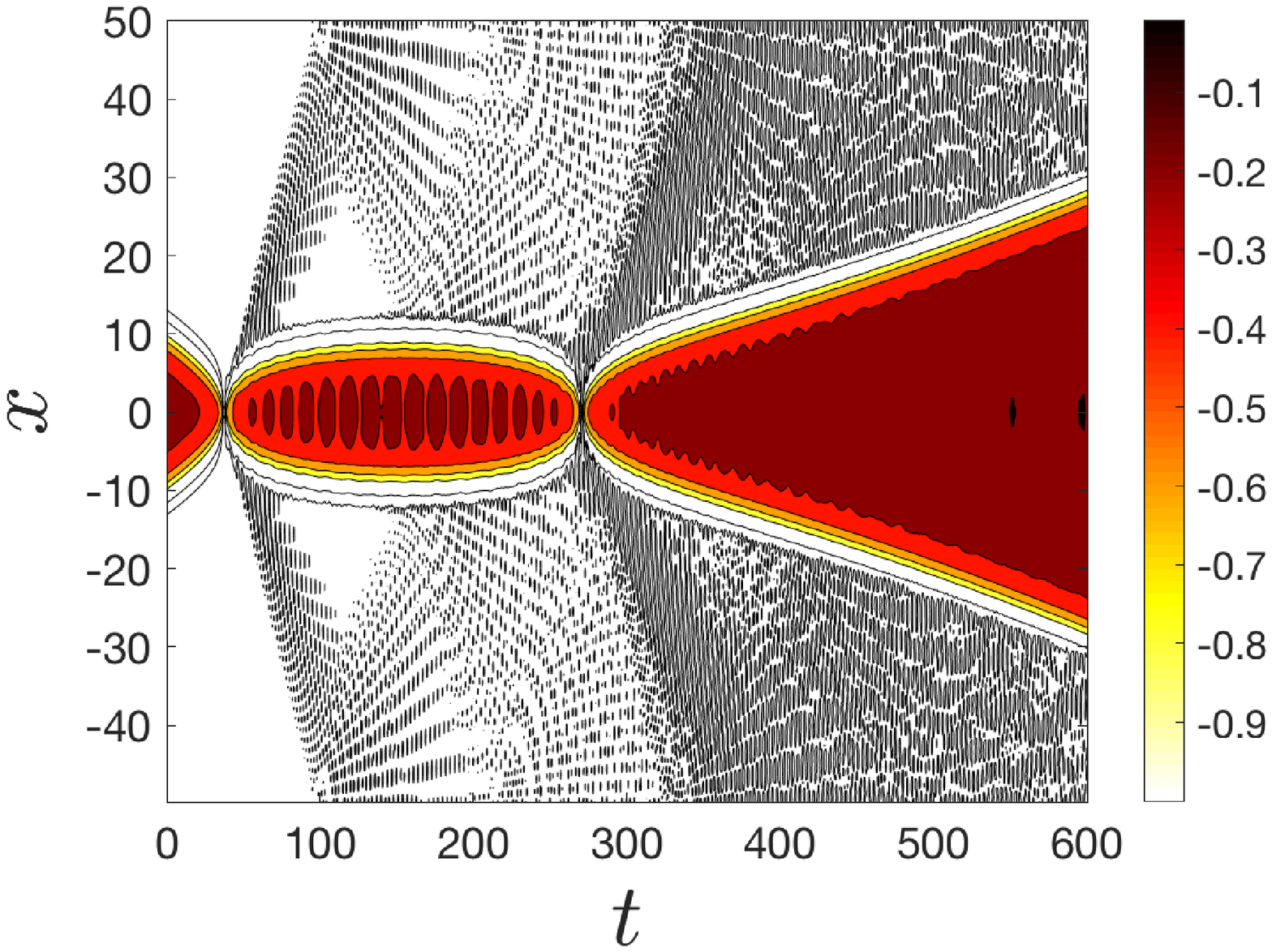}}
\caption{Space-time contour plot of the PDE simulation of the $\phi^8$ kinks collision with (a) damping starting at $x=\pm 30$, and (b) no damping.}
\label{contourDamping}
\end{figure}

The panels in Fig.~\ref{contourDamping}
show the PDE simulations (contour plots) of the $\phi^8$ model for the same incoming velocity; in (a) damping is added, whereas there is no damping for the result shown in (b). In the next section we discuss in detail how initial conditions for such simulations are created. In (a), all radiation created at the time of the first bounce is damped to zero before reaching $\pm 50$, whereas in (b) the radiation from the first bounce is reflected back and reaches the moving kink before the second bounce. In this case, the interfering radiation changes the number of bounces from infinitely many bounces (bound state) to two bounces. Without the added damping, this case would be mistakenly classified as belonging to a two-bounce resonance window. This example, featuring a kink-antikink collision with initial (symmetric) velocities of $0.148673$ and a half-separation of $10$, highlights the need for adding appropriate damping in the simulations.

\section{Creating Initial Conditions With Nonzero Kink Velocity}
\label{sec:Initial_Conditions}

As was shown in Refs.~\cite{our_PRD,our_PRL}, in order to create initial conditions for the case of zero initial velocity (for various separation distances of the kink and antikink), the methods used for the $\phi^{4}$ and other models that exhibit kinks with exponential tails are not appropriate. Instead of a sum or product ansatz, a minimization procedure needs to be applied to any ansatz to obtain the initial position for a kink-antikink combination. The same applies when the initial velocities are not zero (to obtain the initial position) and, additionally, an appropriate initial velocity function must be generated.

In advancing the state of the art beyond kink-antikink interaction with zero initial velocity and towards interactions with nonzero initial velocities, we need to generate initial conditions that represent a moving kink (and a moving antikink), each starting at a specified position, and moving with a specified velocity, at the initial instant of time. We can then check whether our initial conditions are correctly prescribed by running a simulation of the PDE (equation of motion) and tracking the position of each (kink and antikink). From the graphs of their position and velocity, we can ascertain whether the dynamics are ensuing from the intended (specified) initial position and velocity.

To this end, we start by creating initial conditions for a single coherent structure moving at a specified velocity. This structure is, specifically, a traveling wave of the form $u(x,t)=\phi(x-ct)$. From the governing equation of motion~\eqref{eq:EoM}, we obtain an ODE for the shape
with the speed $c$ as a parameter:
%
\begin{equation}
\left(1-c^{2}\right)\phi ^{\prime \prime }(\xi )-V^{\prime }\big(\phi (\xi )\big)=0,
\label{steady1}
\end{equation}
where $\xi=x-ct$ is the moving frame coordinate. 

Implicit solutions to this ODE are known from \cite{Khare.PRE.2014}. Since $u_{t}=-c\phi^\prime(x-ct)$, we have
\begin{equation}
u(x,t=0)=\phi(x) \quad\text{ and }\quad u_{t}(x,t=0)=-c\phi^{\prime}(x)  
\label{initials1}
\end{equation}
for the initial conditions of our dynamic PDE simulations. Here, $t$ subscripts denote partial derivatives with respect to $t$.

As in \cite[Sec.~III.D]{our_PRD} we consider kink ($\phi_{(-1,0);c}$) and antikink ($\phi_{(0,-1);c}$) solutions to Eq.~\eqref{steady1}, with velocity $c$, which connect the minima of $V$ at $-1$ and $0$.
For kink-antikink interactions, we start with a split-domain ansatz \cite[Sec.~III.D]{our_PRD}, stitching together a single kink shifted to the left by the amount $x_0^{}$ and moving to the right with velocity $c$ for $x<0$ and an antikink shifted to the right by an equal amount $x_0^{}$ and moving to the left (velocity $-c$) for $x>0$. Formally the ansatz is given by
\begin{equation}
\varphi(x,t)=[1-H(x)] \varphi_{(-1,0);c}^{}(x+x_0^{})+H(x) \varphi_{(0,-1);-c}^{}(x-x_0^{}).
\label{splitDomain}
\end{equation}
Here, $H(x)$ is the Heaviside unit-step function.

Since this split-domain curve has a ``corner'' at $t=0$ (a discontinuity in the derivative) it is not suitable as an initial condition for the PDE. Instead, it is used as an initializer in a minimization of the norm of the left side of Eq.~\eqref{steady1}, subject to two constraints, $\varphi(-x_0^{})=\tilde{\phi}$ and $\varphi(x_0^{})=\tilde{\phi}$, that keep the locations of the kink and antikink fixed at $-x_0$ and $x_0$ respectively, as was done in our previous works \cite{our_PRD,our_PRL}. 
Specifically we use Eq.~\eqref{splitDomain} as the initializer for a weighted nonlinear least-squares minimization of the objective function 
\begin{equation}\label{eq:min_funct}
    \mathcal{I}[\varphi] = \left\Vert (1-c^2)D_2^{}\varphi-V^\prime(\varphi)\right\Vert_2^2 + C\left|\varphi(-x_0^{})-\tilde{\phi}\right|^2 + C\left|\varphi(x_0^{})-\tilde{\phi}\right|^2,
\end{equation} 
where $\|\cdot\|_2^{}$ is the usual Euclidean norm, $D_2^{}$ is the discrete spectral second derivative matrix, and $C$ is an empirical constant. Then, we can use the minimizer $\varphi_{\min}^{}(x)$ of $\mathcal{I}$ as the initial condition for a direct numerical simulation of kink-antikink collisions, ensuring that our initial condition quantitatively satisfies the PDE to some preset accuracy. We use a weight of $C=500$, which is sufficient to keep the initial kink and antikink locations nearly fixed at $\pm x_0^{}$ during the minimization process. The optimization problem is solved using \textsc{Matlab}'s optimization toolkit, specifically via the {\tt lsqnonlin} subroutine.

Now that we have generated an initial condition $u(x,t=0)$ from the shape function $\phi(x)$ (calculated from the minimized split-domain ansatz), we need to create an initial velocity $u_t(x,t=0)$ that sends the kink and antikink towards each other at speeds $c$. Since we are no longer working with a sum or product ansatz, we need to create the initial velocity function directly from the shape function $\phi(x)$. As for the single kink case, $-c\phi^\prime(x)$ would seem like a reasonable choice, except that this would send both the kink and antikink in the same direction, rather than at each other. Therefore, we choose
\begin{equation}
u_t(x,t=0)= c\:\sgn(x)\:\phi^{\prime }(x)  \label{Eq 3}
\end{equation}
for the initial velocity function.

\begin{figure}[t!]
\centering
\subfigure[]{\includegraphics[width=0.46\textwidth]{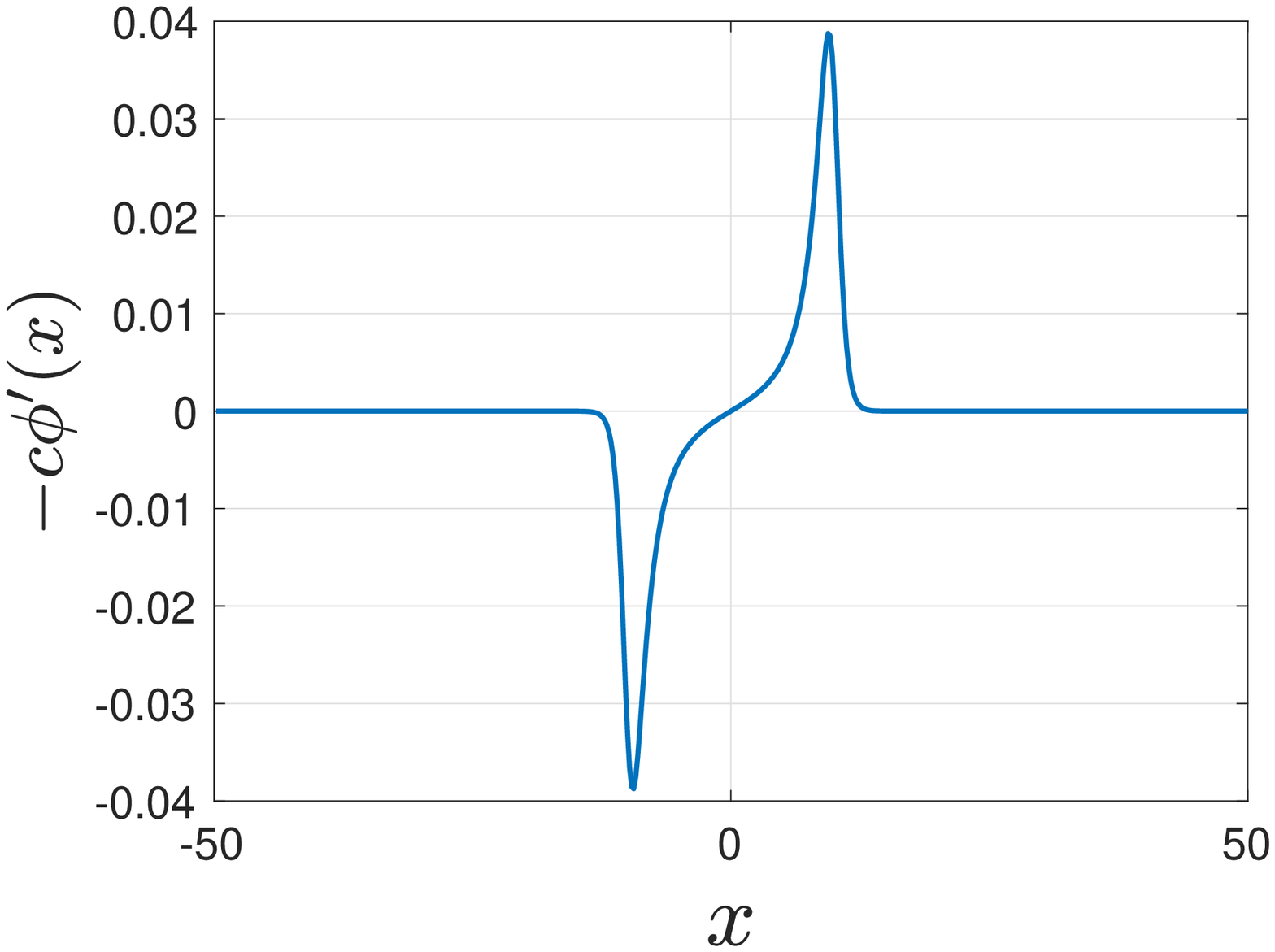}}%
\subfigure[]{\includegraphics[width=0.46\textwidth]{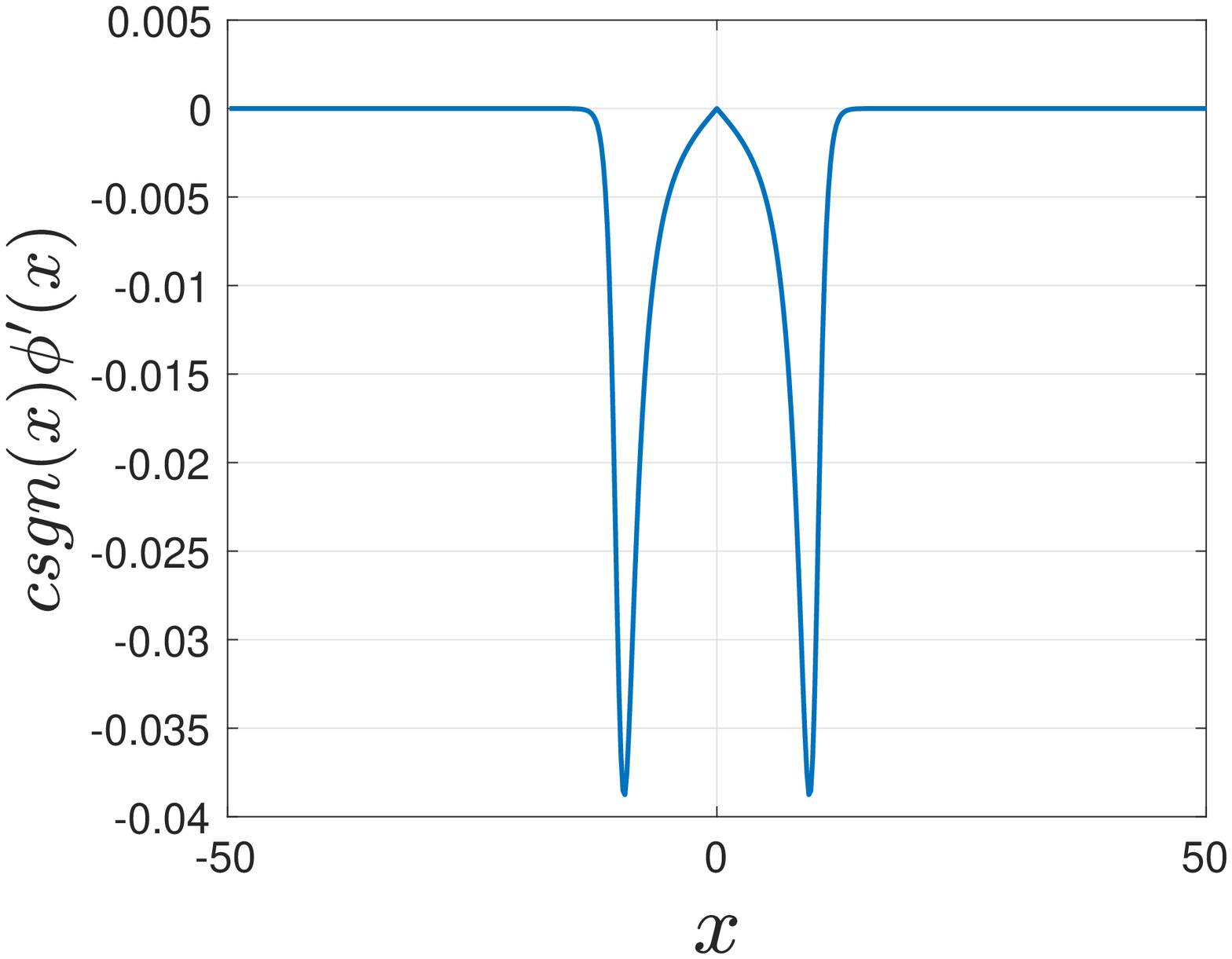}}
\caption{(a) Graph of $-c\phi^\prime(x)$ for the kinks of the $\phi^8$ model. (b) Graph of $c\sgn(x)\phi^\prime(x)$.}
\label{initialVel}
\end{figure}

Figure \ref{initialVel}
shows graphs of the functions $-c\phi^\prime(x)$ and $c\sgn(x)\phi^\prime(x)$ for the $\phi^8$ kink-antikink configuration. While $-c\phi^\prime(x)$ is smooth, $c\sgn(x) \phi^\prime(x)$ has a ``corner'' (discontinuity in the derivative of the velocity field, i.e., in the acceleration) at $t=0$. Thus, after minimization, we have eliminated the corner in $u(x,t=0)=\phi (x)$, but there is now one in $u_{t}(x,t=0)=c\sgn(x)\phi^\prime(x)$. Next, we assess how ``well'' these initial conditions work in a dynamics simulation.

\begin{figure}[t!]
\centering
\subfigure[]{\includegraphics[width=0.46\textwidth]{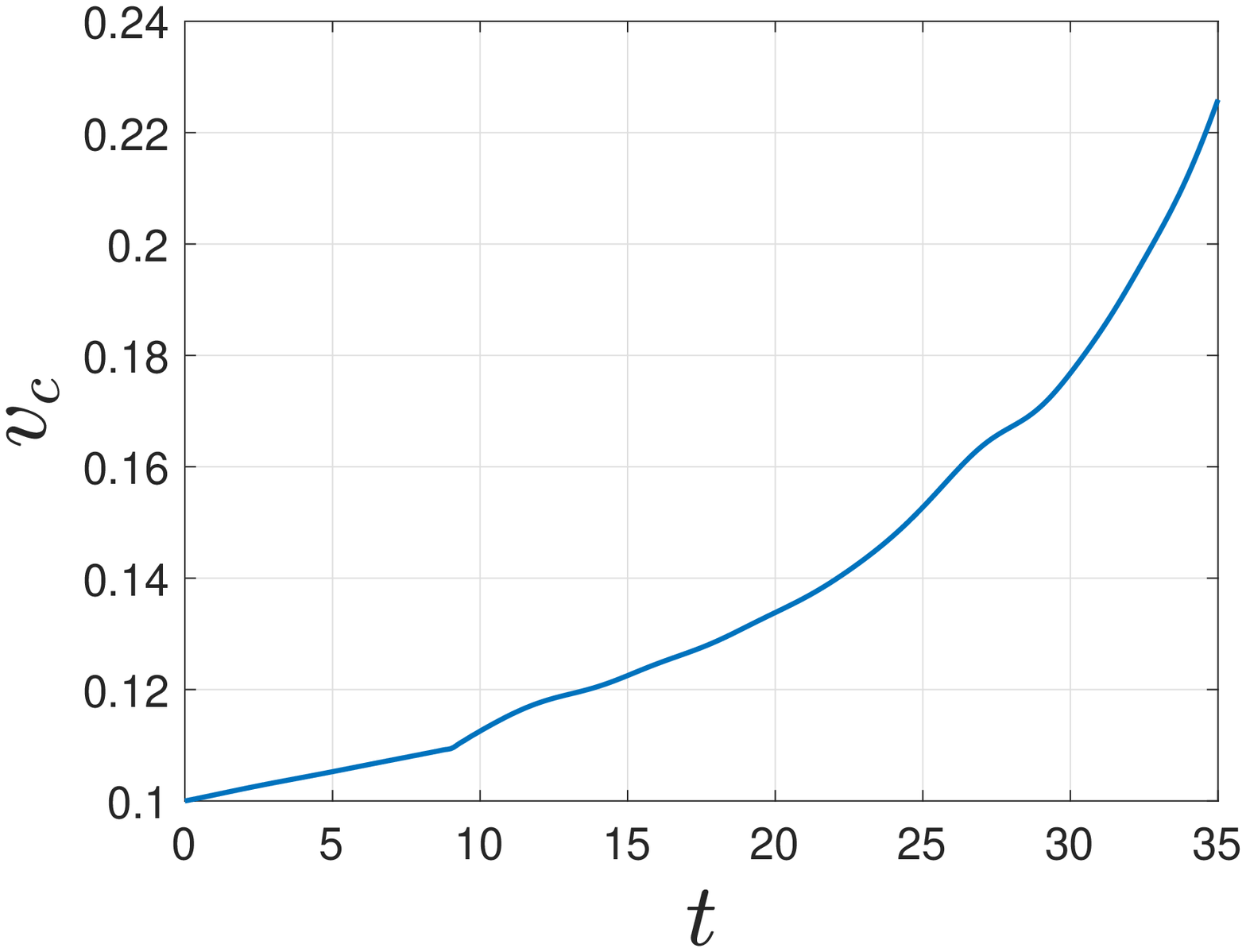}}
\subfigure[]{\includegraphics[width=0.46\textwidth]{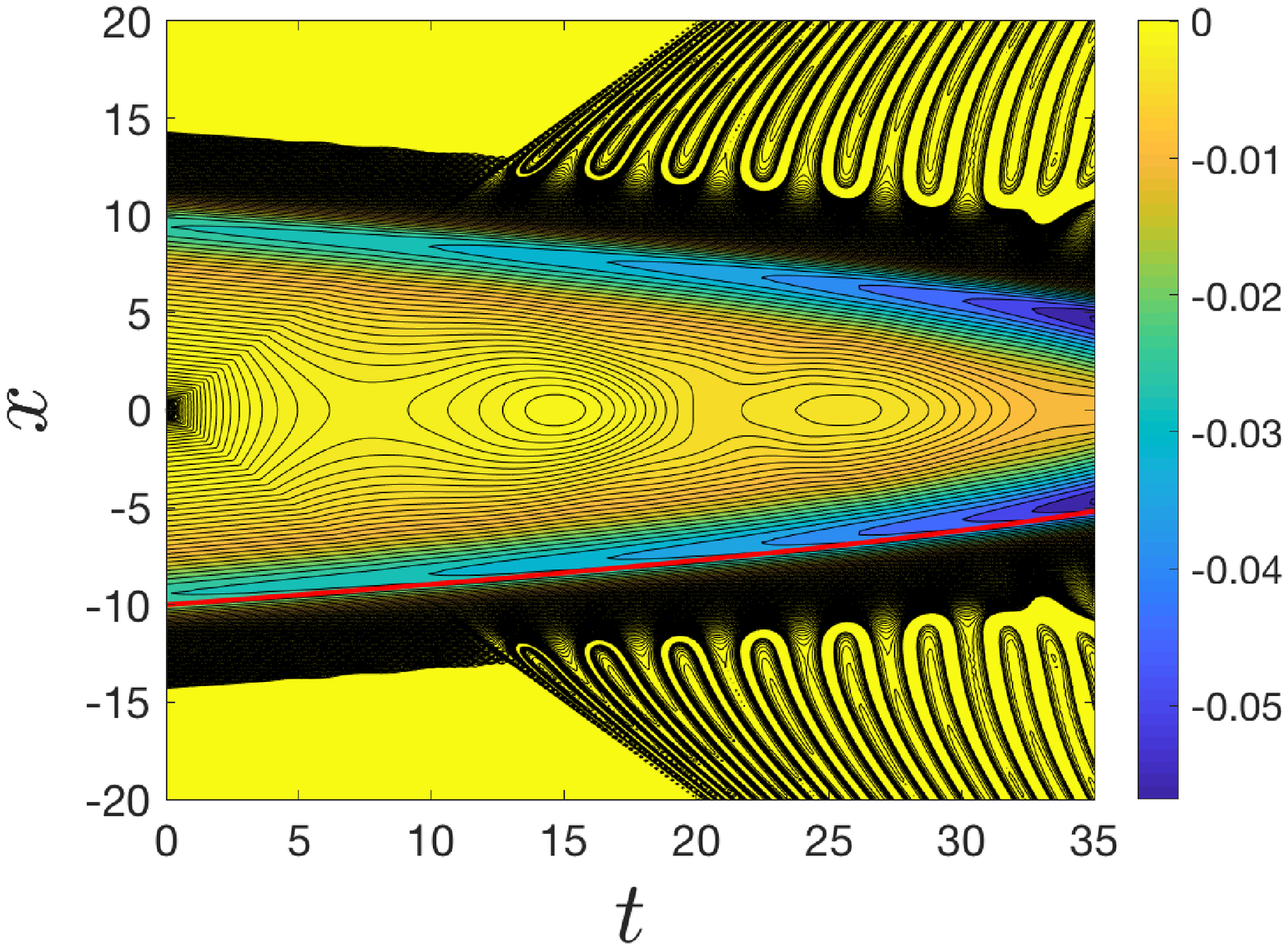}}
\caption{Initial kink half-separation $x_0^{}=10$, initial kink velocity $c=0.1$, using $u_t(x,t=0)$ as in Fig.~\protect\ref{initialVel}(b) for the $\phi^8$ model. (a) Evolution
of the kink velocity function over time. (b) Space-time contour plot of $u_t(x,t)$.}
\label{velAndContourOriginal}
\end{figure}

The effect of the discontinuity (in the acceleration field) becomes apparent in Fig.~\ref{velAndContourOriginal}. Figure \ref{velAndContourOriginal}(a) shows ``bumps'' in the velocity function of the kink at around $t=10$ and $t=27$. Figure \ref{velAndContourOriginal}(b) is a contour plot of $u_{t}(x,t)$ (with carefully chosen contour levels to emphasize the effect being described next). One can observe the 
resulting small amplitude waves in the $u_{t}(x,t)$ field after $t=0$ propagating in both
directions, and intersecting the path of the kink (in red) and antikink at about $t=10$, causing the bump in the velocity function.
The goal of finding ``good'' initial conditions for a given initial kink velocity is to simulate kink-antikink collisions, with the initial conditions given at 
any prescribed separation.  This is relatively easy with the $\phi^4$ model, since tails are exponential at both ends and the sum ansatz with a relatively small separation is sufficient because the kinks' overlap is exponentially small in this case.

To understand what a ``correct'' $u_{t}(x,0)$ should look like in the more subtle case considered herein, we propose the following approach. We assume that $u_{t}(x,0)=0$ would give an \textquotedblleft ideal\textquotedblright\ initial velocity. We start with a half-separation of $x_0^{}=30$ units and zero initial velocity, and let the simulation run to the point at which the half-separation is (very close to) $10$ units. Then we restart the simulation in three different ways.

\begin{enumerate}
\item The first way is to restart the simulation with the field configuration, i.e., $u$ and $u_{t}$, obtained at the end of the first simulation. These should represent an ``ideal" set of initial conditions as they are simply continuing the run starting at half-separation of 30 and a zero velocity. From Fig.~\ref{velAndContourPerfect}, we see that both plots are smooth, as expected. No irregularities are visible in either (a) the kink velocity plot or (b) the space-time contour plot of $u_t(x,t)$. Also, there is no corner in the $u_{t}(x,0)$ plot, thus there is no spurious propagation of small-amplitude radiation/waves. This is the comparison case, representing what we believe we should observe if the correct initial conditions are used.

\begin{figure}[t!]
\centering
\subfigure[]{\includegraphics[width=0.46\textwidth]{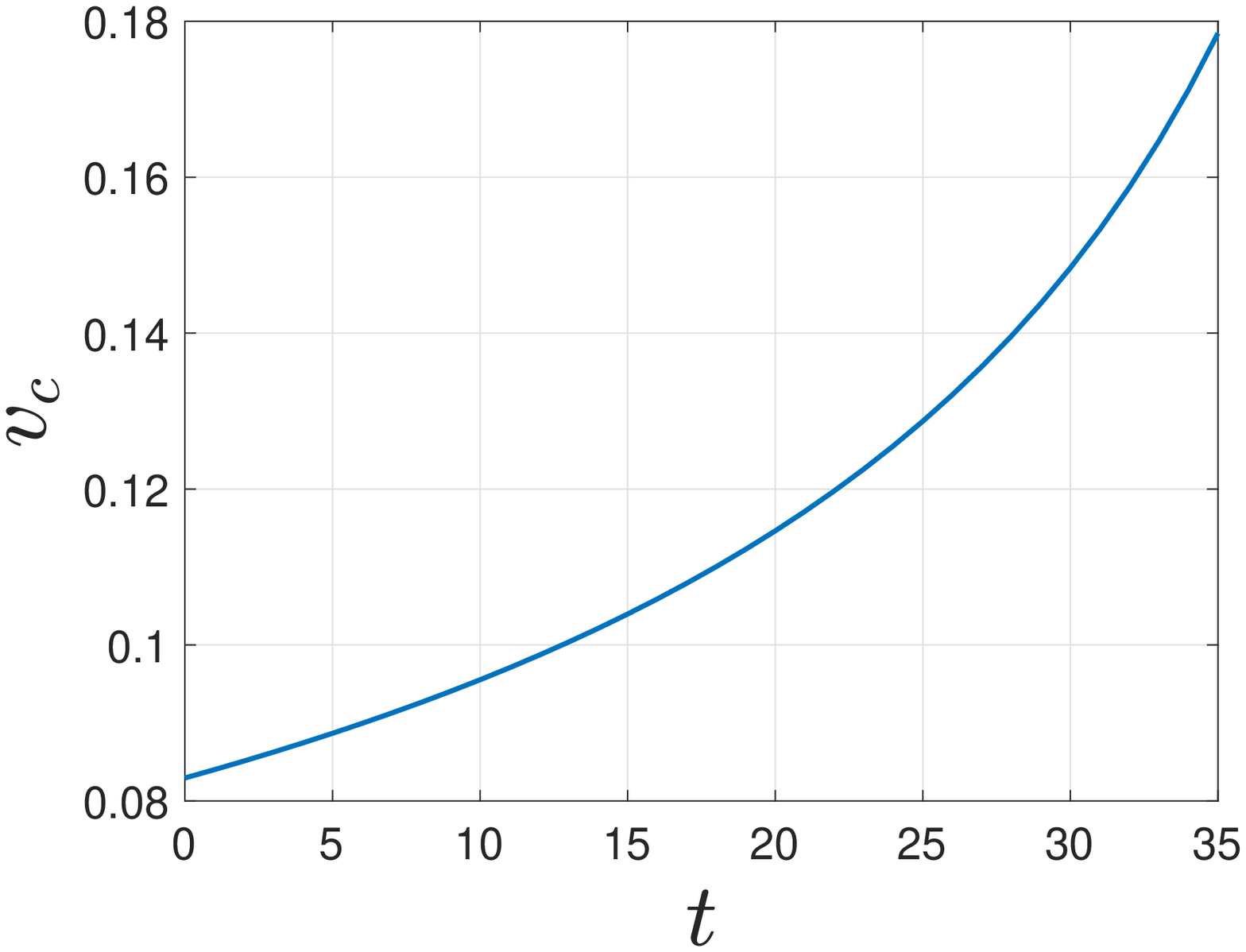}}%
\subfigure[]{\includegraphics[width=0.46\textwidth]{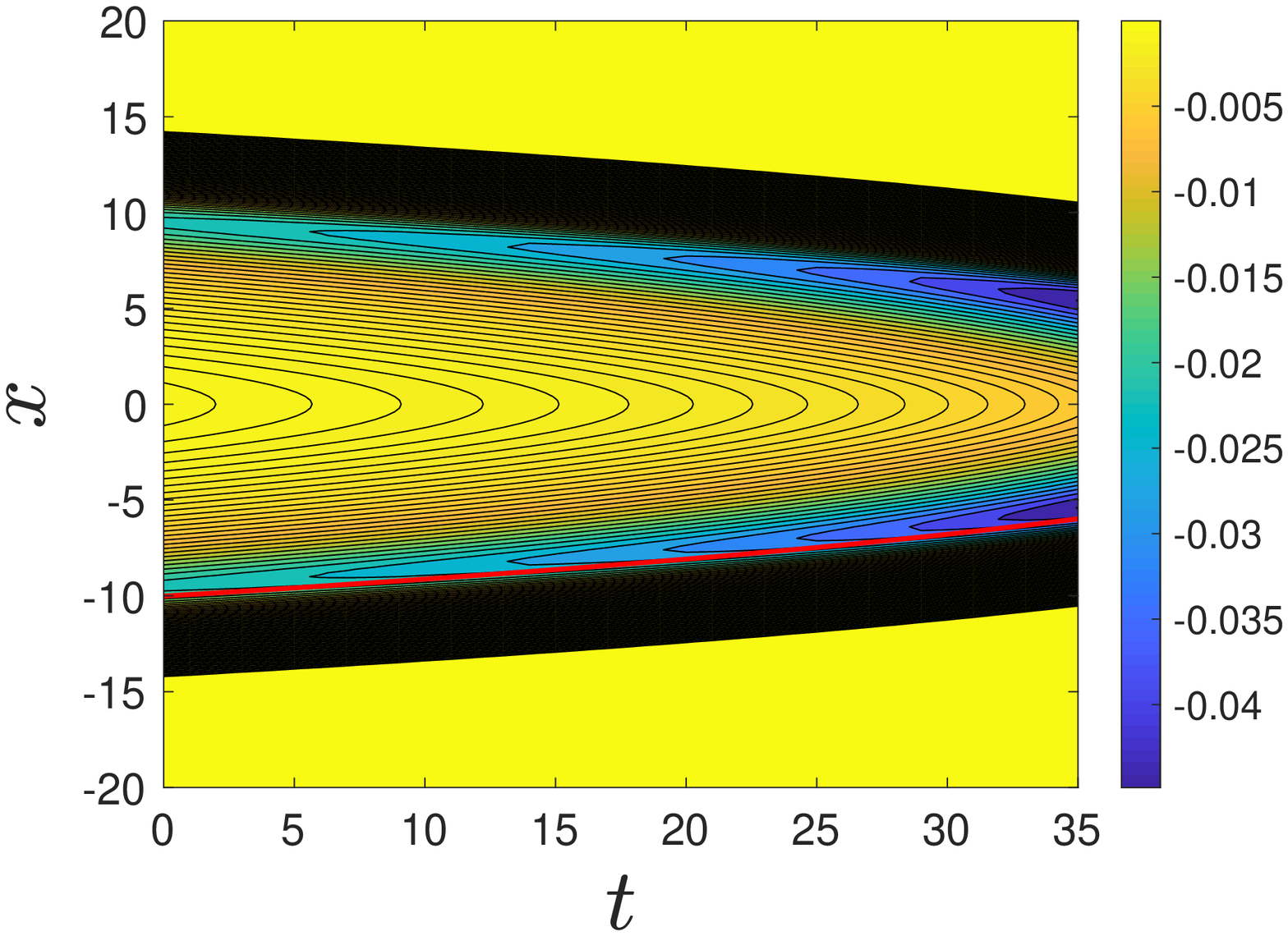}}
\caption{Ideal initial conditions by restarting PDE from $x_0^{}\approx 10$ after initial run with half-separation $x_0^{}=30$ and initial velocity zero for the $\phi^8$ model. (a) Kink velocity function. (b) Space-time contour plot of $u_t(x,t)$.}
\label{velAndContourPerfect}
\end{figure}

\item The second way is to restart the PDE simulation with the ``standard'' initial conditions. This approach uses a minimized $u(x,0)$, which uses the kink separation found at the end of the original run, and $u_t(x,0)=v_{\rm in}^{}\sgn(x)u_x(x,0)$, where $v_{\rm in}^{}$ is the final velocity at the end of the original run. As a consequence, a corner is created in the $u_t(x,0)$ plot, and propagating small amplitude
waves are observed in the $u_{t}(x,t)$ contour plot, as discussed above.
In Fig.~\ref{velAndContourOriginal2}(a), we show the kink velocity for the ideal case alongside the one for the current case; one can easily see the difference between the two. Fig. 6(b) shows the $u_t(x,t)$ contour plot.

\begin{figure}[h]
\centering
\subfigure[]{\includegraphics[width=0.46\textwidth]{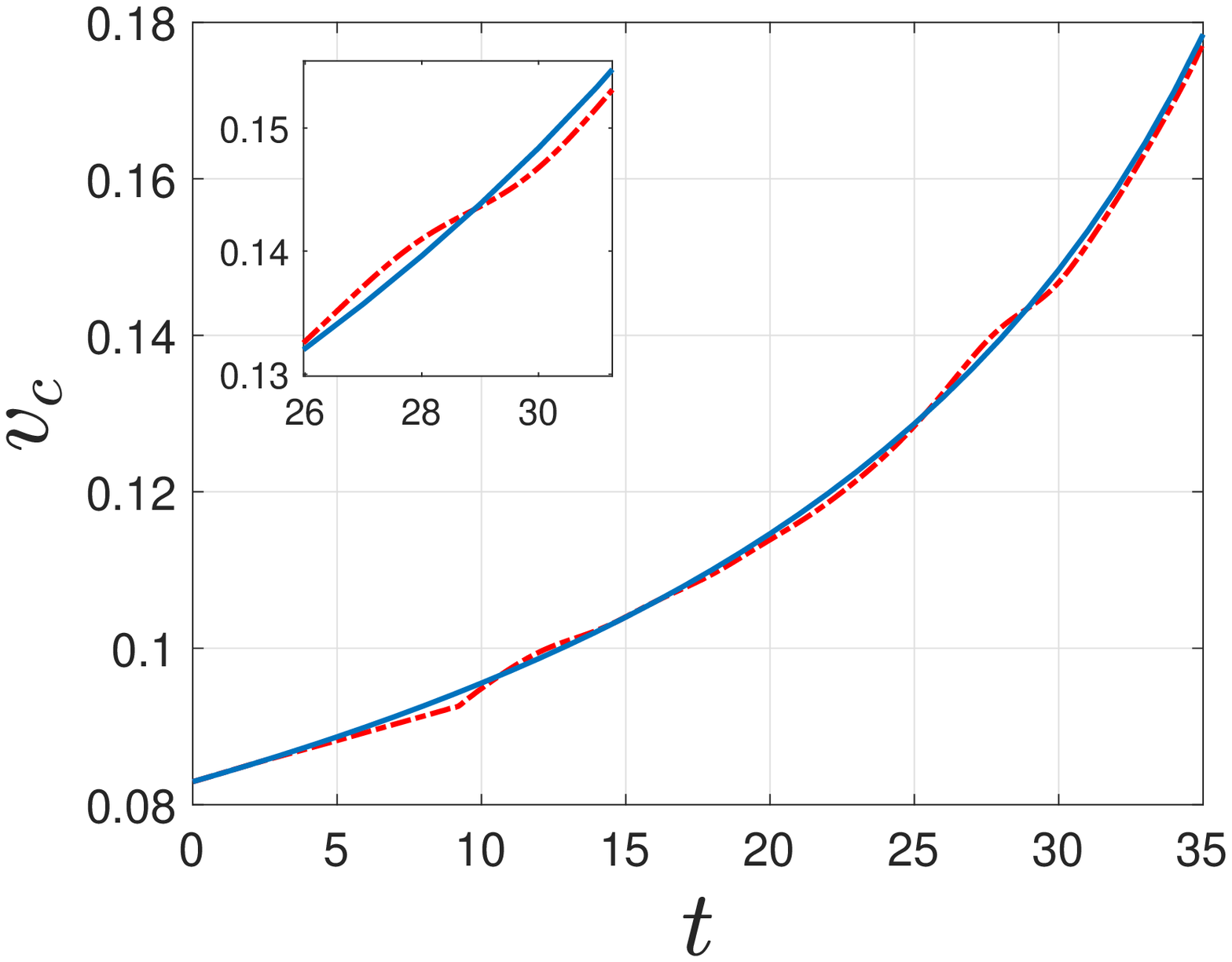}}%
\subfigure[]{\includegraphics[width=0.46\textwidth]{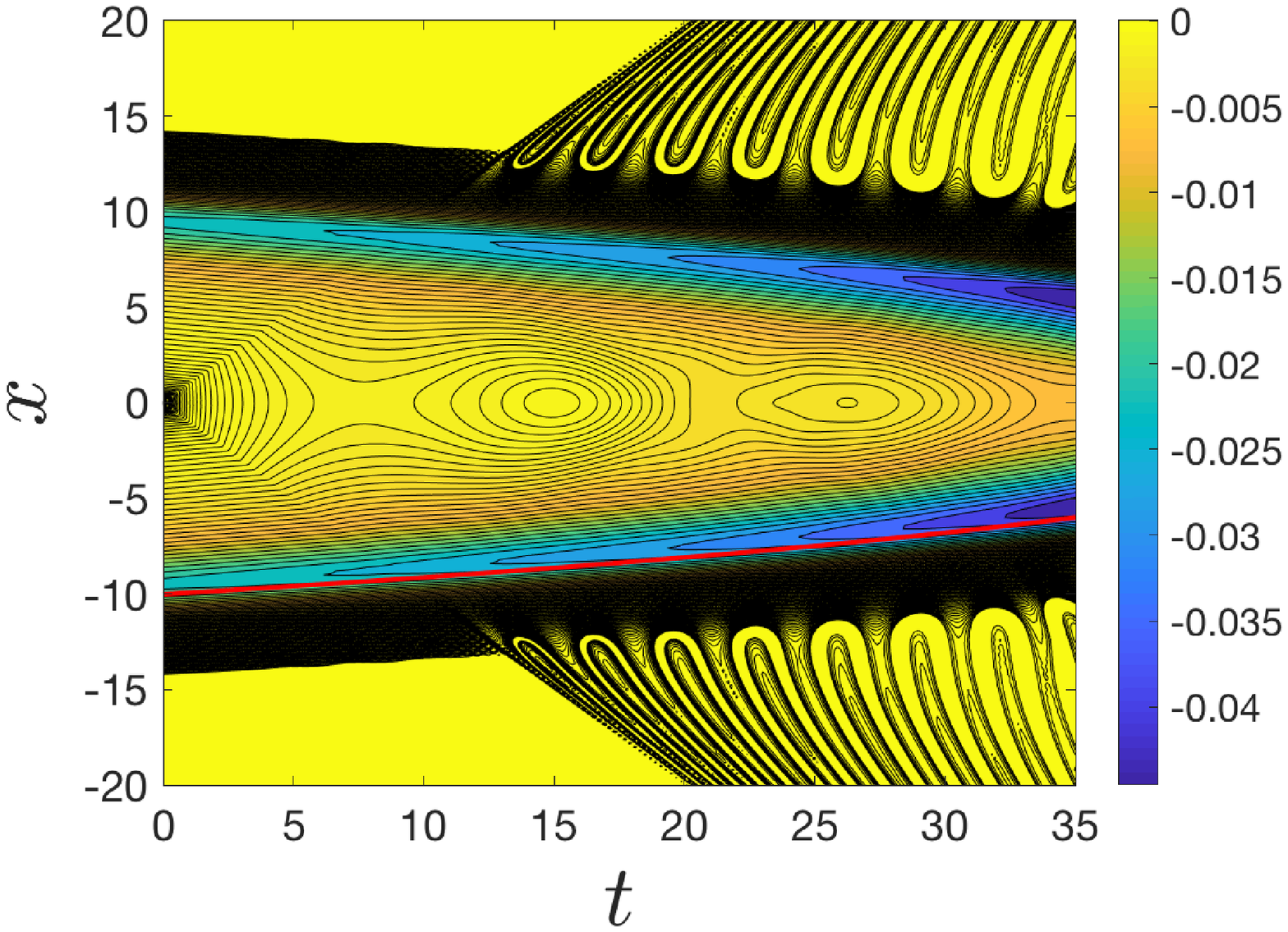}}
\caption{Initial conditions using minimized initial position, and initial velocity given by $v_{\rm in}^{}\sgn(x) u_x(x,0)$ as shown in Fig.~\ref{initialVel}(b) for the $\phi^8$ model. (a) Kink velocity function (red/dash-dotted curve) with ideal kink velocity from Fig.~\ref{velAndContourPerfect} (solid/blue curve); inset shows $v_{\rm c}$ versus $t$ at late times. (b) Space-time contour plot of $u_t(x,t)$.}
\label{velAndContourOriginal2}
\end{figure}

\item The third way is to use a process we dub ``second minimization'', where we
take the minimized field configuration $u(x,0)$, and then minimize the norm of the PDE again (subject to some constraints discussed shortly) to obtain $u_t(x,0).$ Here, by ``minimize again" we mean that the search algorithm fixes $u(x,0)$ (the minimized $u(x,0)$ from the previous case) and varies only $u_{t}(x,0)$ to find a minimum for the 2-norm of the PDE given in Eq. (2) plus constraints. This requires integrating Eq.~\eqref{eq:EoM} over some range of $t$ values, using the initial position determined from the first minimization, for each iteration of the second minimization process. We integrate over the interval $[0, 0.04]$ and then use the values of $u$ at $t=0, 0.02, 0.04$ when applying the constraints (thus we calculate two new $t$ values). This integration range and choice of $t$ values was chosen in order to balance computing time and accuracy. The constraints that we add are to keep the starting kink velocity at the intended $c$ value (in this case determined by the end of the first run in which $x_0^{}=30$ and the initial velocity is zero, but can be user-specified when using second minimization in general). The constraints that we use are $u_t=-c u_x$ in the interval $-x_0^{}-2\le x \le -x_0^{}+1.6$ and $u_t=c u_{x}$ in the interval $x_0^{}-1.6\le x \le x_0^{}+2$, both applied only at $t=0$. Thus the objective function for 
the second minimization is

\begin{equation}
    \mathcal{J}[u] = \left\Vert D_t u_t-D_2^{} u+V^\prime(u)\right\Vert_2^2 + C\left\Vert u_{t}+c u_{x}\right\Vert_2^2 + C\left\Vert u_{t}-c u_{x}\right\Vert_2^2,
\end{equation} 
where in the first term $x$ ranges over the all $x$ and over the $t$ values $0$, $0.02$, $0.04$. $D_t$ represents differentiation with respect to $t$ using first-order finite differences, and $u_t$ is produced by numerical integration over the interval [0,0.04] as previously described. Thus $D_t u_t$ gives a numerical approximation to $u_{tt}$. In the second and third terms $x$ has the ranges just given, but we use only $t=0$. As for the first minimization, the optimization problem is solved using \textsc{Matlab}'s {\tt lsqnonlin} subroutine. Figure \ref{velAndContourImproved}(a) shows the initial velocity plot after the second minimization showing that the corner has been smoothed (compared to non-minimized initial velocity in Fig.~\ref{initialVel}(b)).
\end{enumerate}

Figure \ref{velAndContourImproved}(b) shows the kink velocity for the ideal case alongside the one for the current case; there is no noticeable variation between the two, at this scale. Figure \ref{velAndContourImproved}(c)  shows the velocity contour plot, which is very close to the ideal velocity contour plot from Fig.~\ref{velAndContourPerfect}(b). A drawback to minimizing a second time is that the second minimization requires simulating the evolution of the PDE over some time interval, for each iteration of the minimization procedure. This step increases the computation time required to create the initial conditions by about an order of magnitude, compared to the case in which only the initial position function undergoes minimization. Thus, while this procedure is more accurate in terms of its proposed initial waveform, it is also more computationally
costly.

\begin{figure}[t!]
\centering
\subfigure[]{\includegraphics[width=0.46\textwidth]{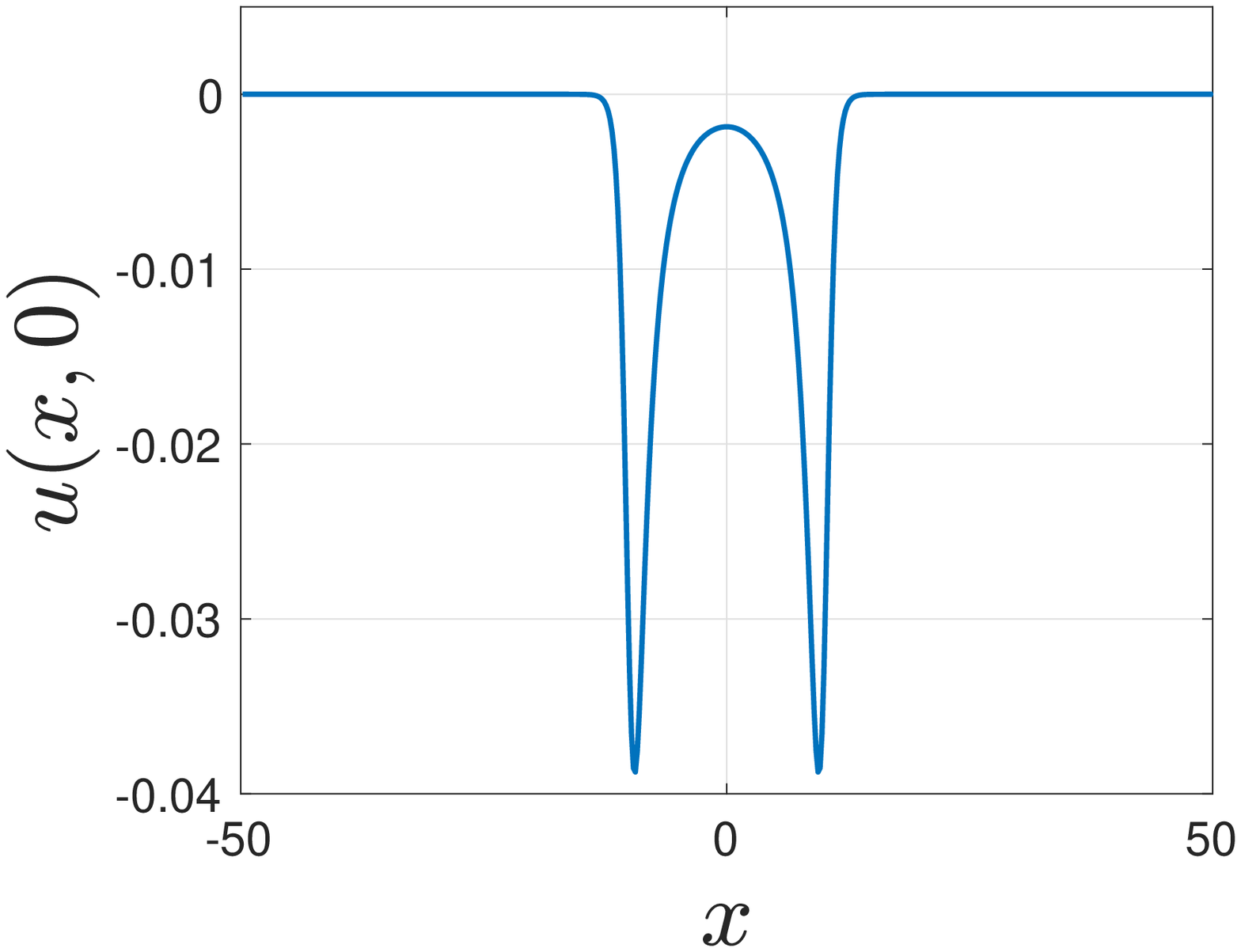}}%
\subfigure[]{\includegraphics[width=0.46\textwidth]{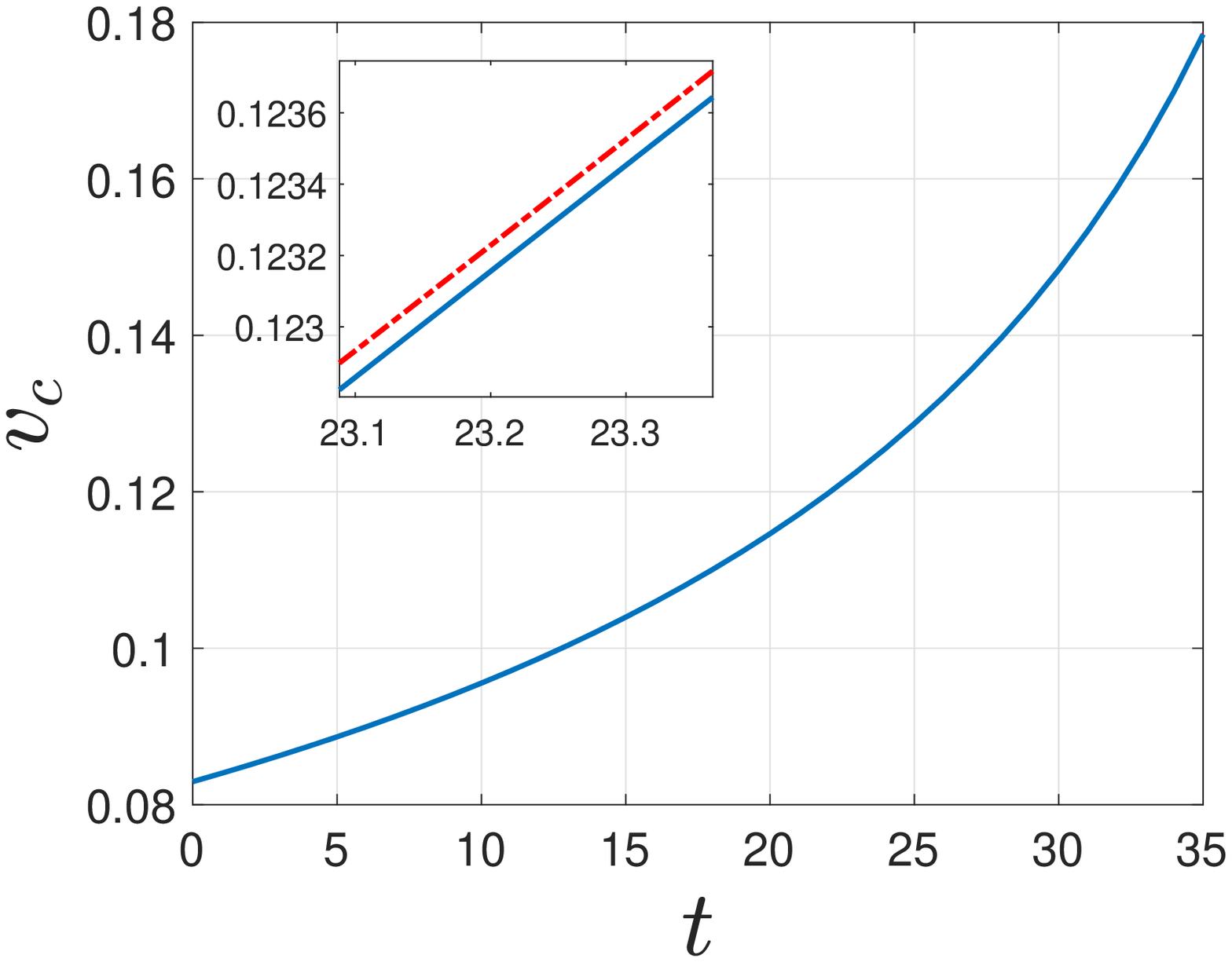}}
\subfigure[]{\includegraphics[width=0.46\textwidth]{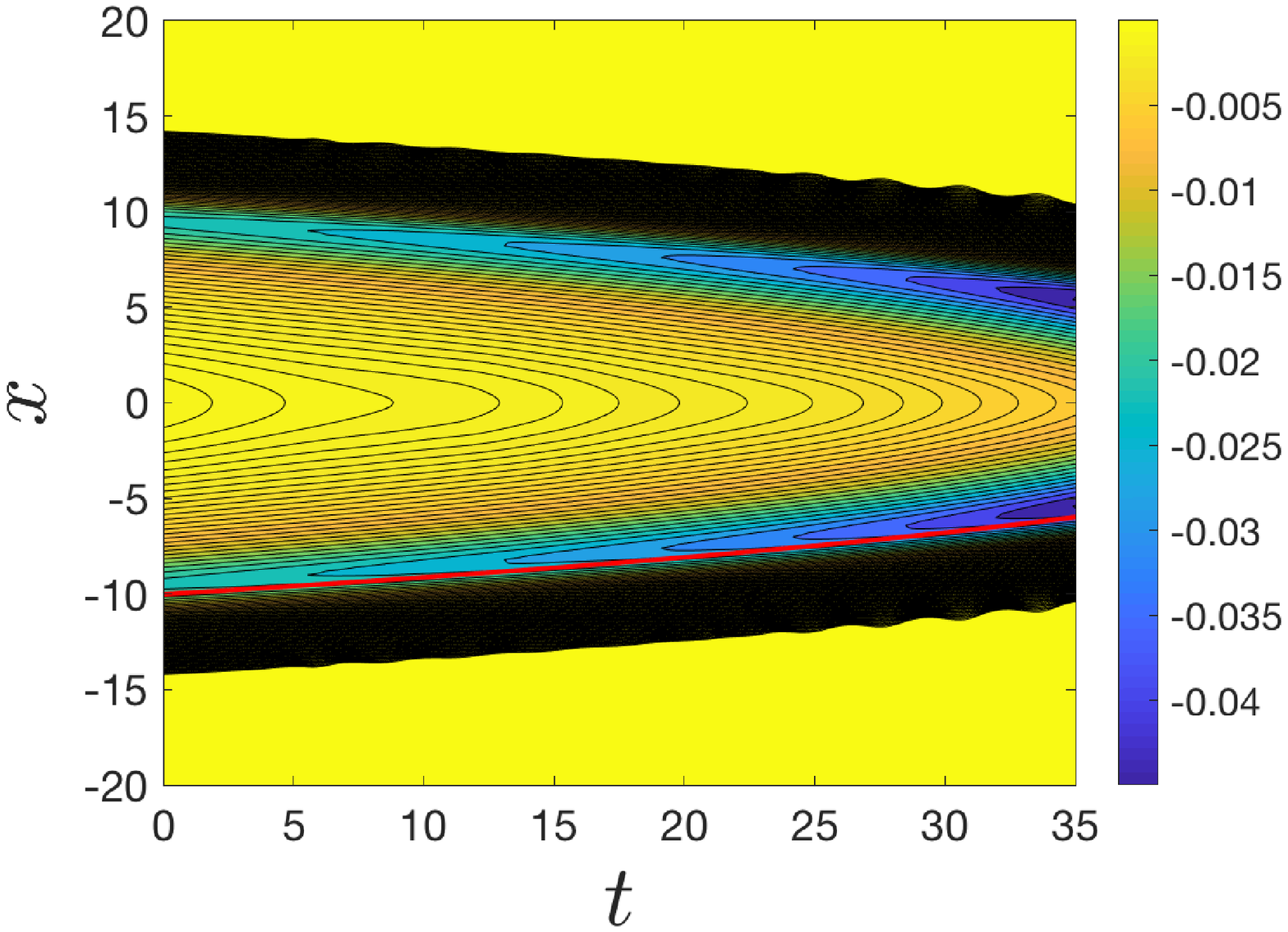}}
\caption{Initial conditions using minimized initial position, and (subsequently) minimized initial velocity for the $\phi^8$ model; see the relevant explanation in the text. (a) Minimized initial velocity plot. (b) Kink velocity function (red dash-dotted curve) and ideal kink velocity function (blue solid curve); inset shows $v_{\rm c}$ versus $t$ at late times. (c) Space-time contour plot of $u_t(x,t)$.}
\label{velAndContourImproved}
\end{figure}

\section{The Relationship between Velocity-out and Velocity-in Upon Collision}
\label{sec:V_in-V_out}

In the previous section, we showed that initial conditions generated by first minimizing a split-domain ansatz for the initial field configuration $u(x,t=0)$ of the kink-antikink combination and then minimizing again to obtain an improved initial velocity $u_{t}(x,t=0)$ yields the ``best'' initial conditions for analyzing kink-antikink collisions via numerical simulation of the PDE. However, there is a cost: the computational time required to create many such initial conditions makes it less feasible to run a sufficient number of PDE simulations to generate accurate velocity-out ($v_{\rm out}^{}$) versus velocity-in ($v_{\rm in}^{}$) plots. Fortunately, we have found that the $v_{\rm in}^{}$ -- $v_{\rm out}^{}$ relationships predicted by the once-minimized initial conditions (no $u_{t}$ minimization) and the corresponding relationship obtained via the twice-minimized initial conditions have a very similar structure, which we will elaborate on further towards the end of this section. Therefore, we now discuss the numerical results for the once-minimized case.

\subsection{$\phi^8$ case}

\begin{figure}[t!]
\centering
\subfigure[]{{\includegraphics[width=0.49\textwidth]{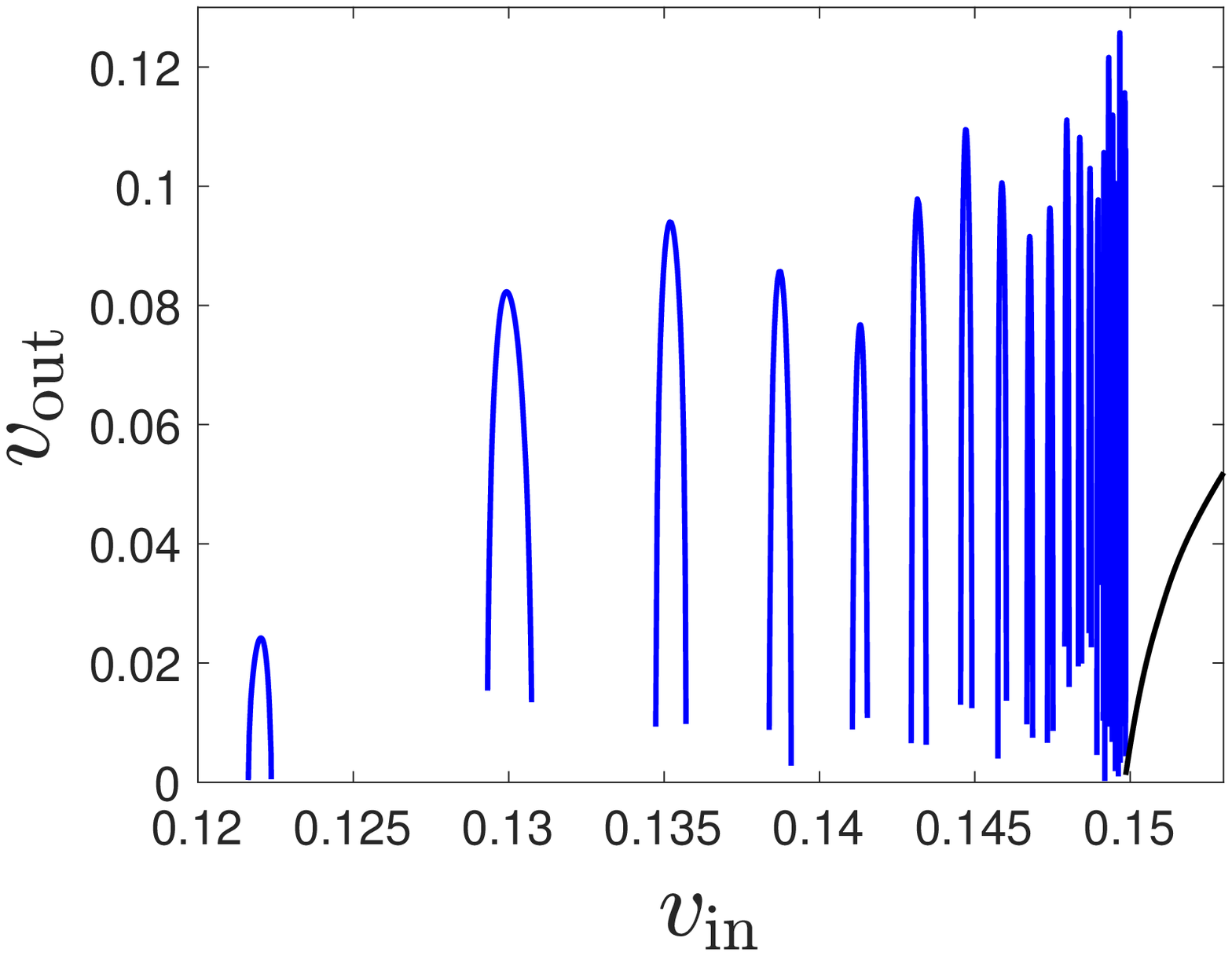}}} \subfigure[]{{\includegraphics[width=0.49\textwidth]{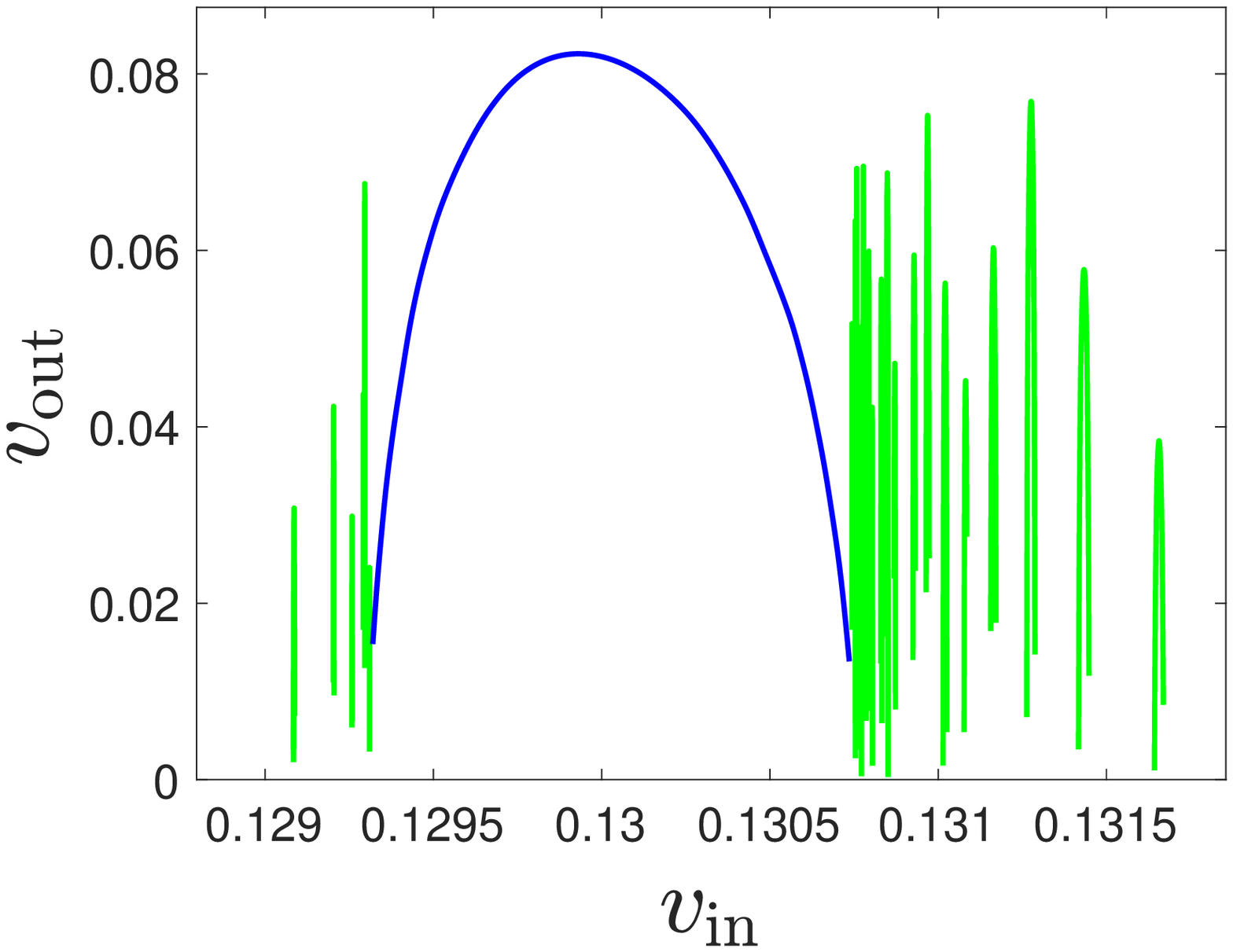}}}
\subfigure[]{{\includegraphics[width=0.49\textwidth]{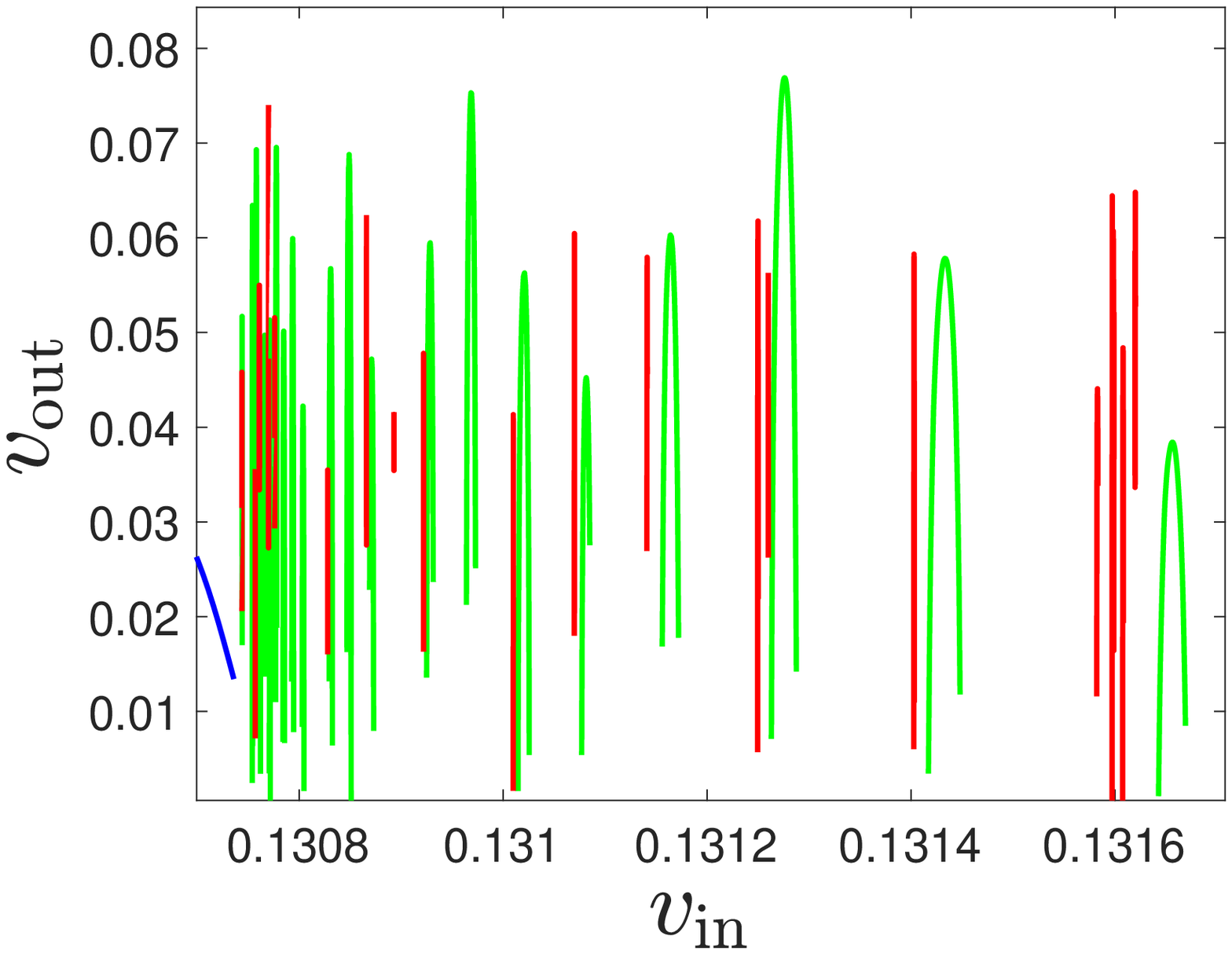}}}
\subfigure[]{{\includegraphics[width=0.49\textwidth]{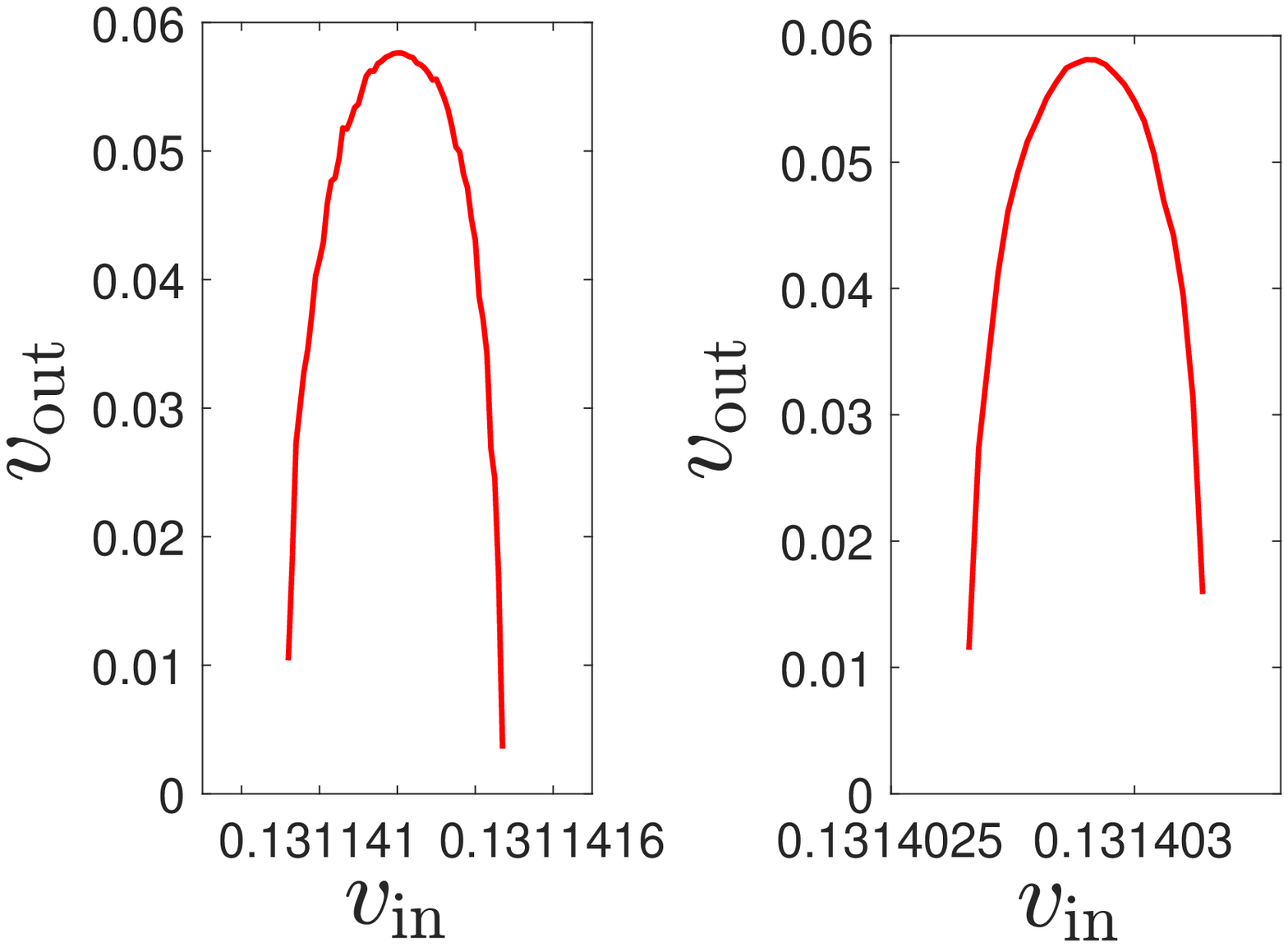}}}
\caption{Plots of the escape velocity $v_{\rm out}^{}$ as a function of the initial velocity $v_{\rm in}^{}$ for the $\phi^8$ model. (a) The two-bounce resonance windows (blue) and one-bounce window (black) are shown; the critical velocity is $v_{\rm cr}^{}\approx 0.1499$.  (b) Zoomed portion of (a) about the second from the left two-bounce window  with three-bounce windows (green) on its edges. (c) Zoomed portion of the right edge of the two bounce window in (b) with added four-bounce windows (red). (d) Zoomed portion of (c) about two 
typical four-bounce windows. }
\label{vin_vout-8}
\end{figure}

In Fig.~\ref{vin_vout-8}, 
we show the $v_{\rm in}^{}$ -- $v_{\rm out}^{}$ curves for the chosen example $\phi^{8}$ model. Panel (a) shows the first several two-bounce windows (in blue) as they accumulate at the critical velocity $v_{\rm cr}^{}=0.1499$. 
Black color represents one-bounce values which occur beyond the critical value $v_{\rm cr}^{}$.
On either side of each two-bounce window one can see
in panel (b) three-bounce windows (in green) which accumulate at the boundaries of the two-bounce windows. Indeed, the panel shows a zoomed-in version of this for the two-bounce window centered at about $v_{\rm in}^{}=0.13$. This fractal-like structure appears to continue for four-bounce windows at the edges of the three-bounce window and so on, but it becomes very difficult to calculate the four-bounce and higher-bounce windows to any reasonable degree of accuracy. Four-bounce values of $(v_{\rm in}^{},v_{\rm out}^{})$ in Fig.~\ref{vin_vout-8}(c) appear as points (red dots). 
Panel (d), in turn, shows some prototypical examples 
(among the many similar ones in panel (c)) of four-bounce
windows.

Figure \ref{vin_vout-8} shows many similarities with published $v_{\rm in}^{}$ -- $v_{\rm out}^{}$ graphs for the $\phi^4$ model, such as a self-similar structure \cite{Anninos.PRD.1991}, as well as some differences. In particular, the maximum values of $v_{\rm out}^{}$ for the two-bounce windows linearly increase for the $\phi^4$ model, whereas for the $\phi^8$ model the maximum values, while increasing overall, also show a pattern of alternately increasing and decreasing.

\begin{figure}[t!]
\centering
\subfigure[]{{\includegraphics[width=0.49\textwidth]{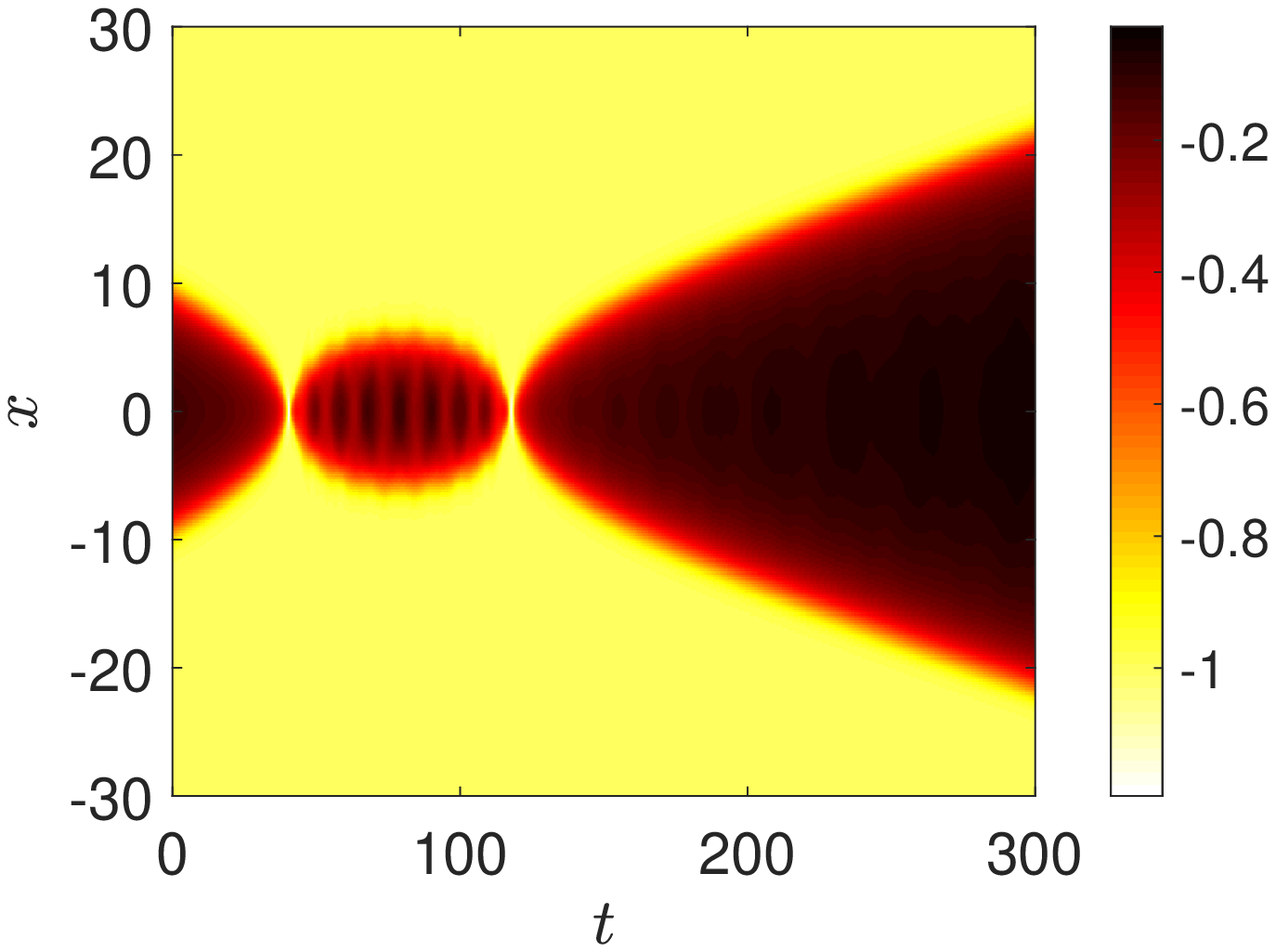}}} \subfigure[]{{\includegraphics[width=0.49\textwidth]{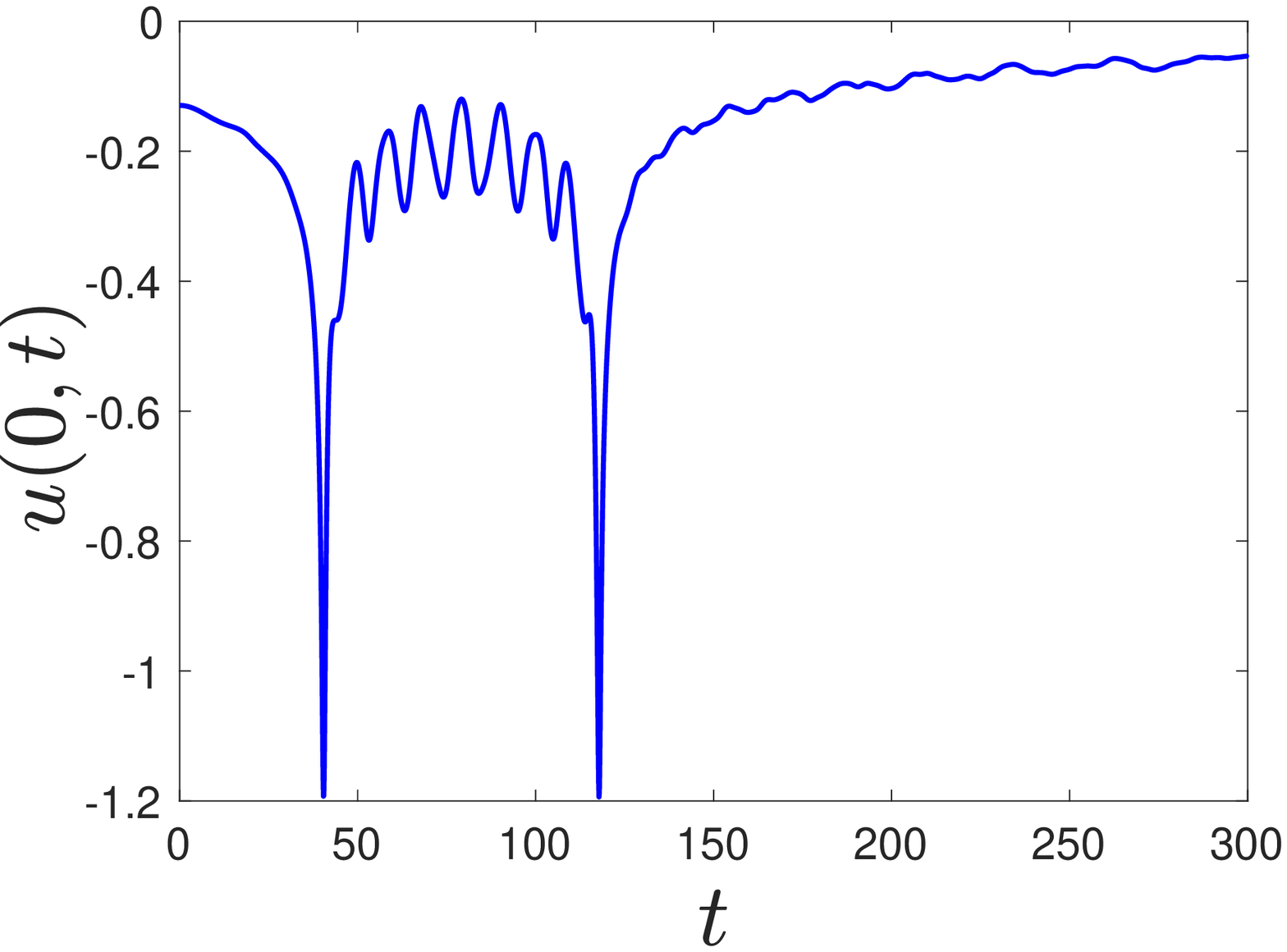}}}
\subfigure[]{{\includegraphics[width=0.49\textwidth]{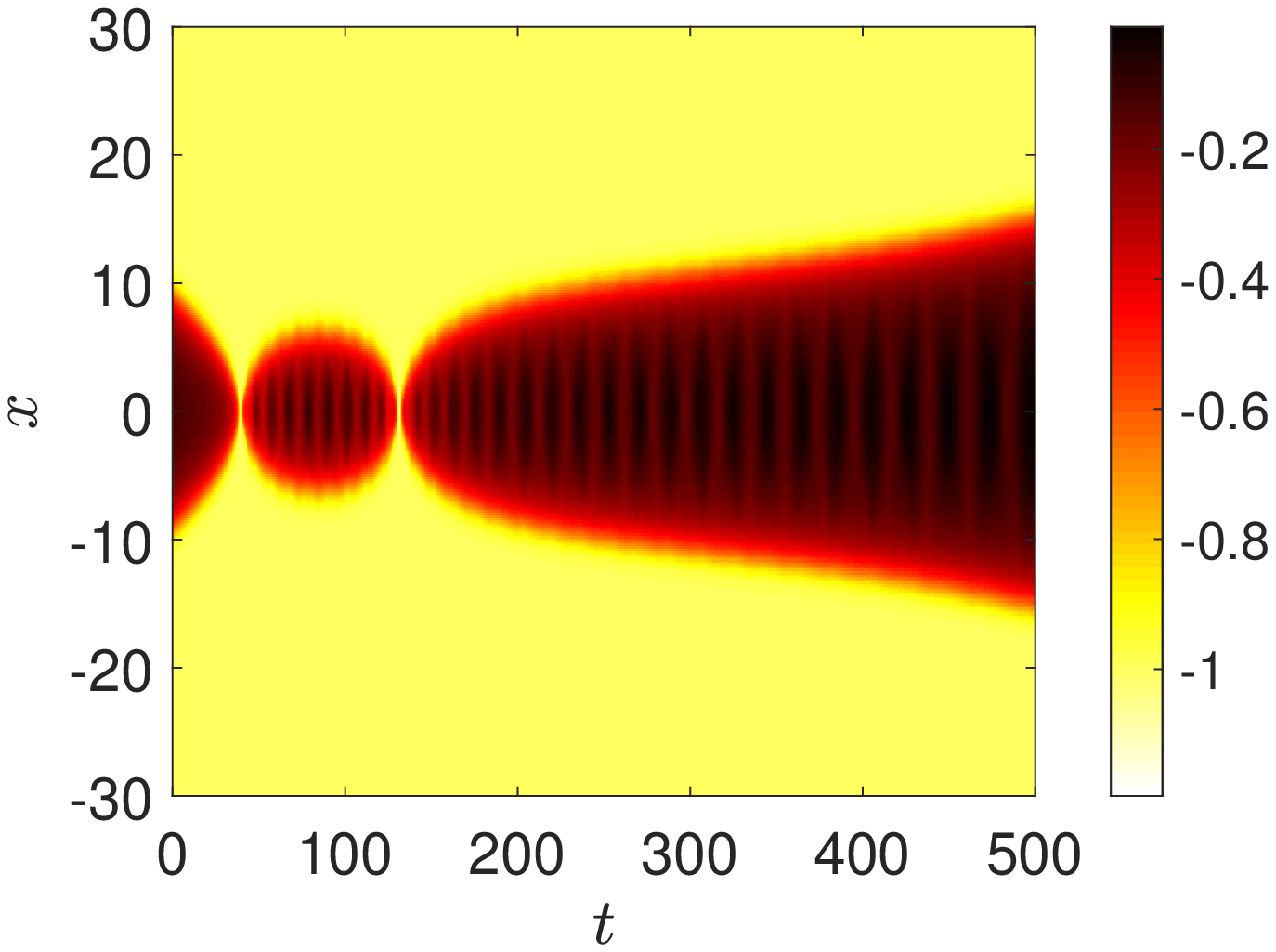}}} \subfigure[]{{\includegraphics[width=0.49\textwidth]{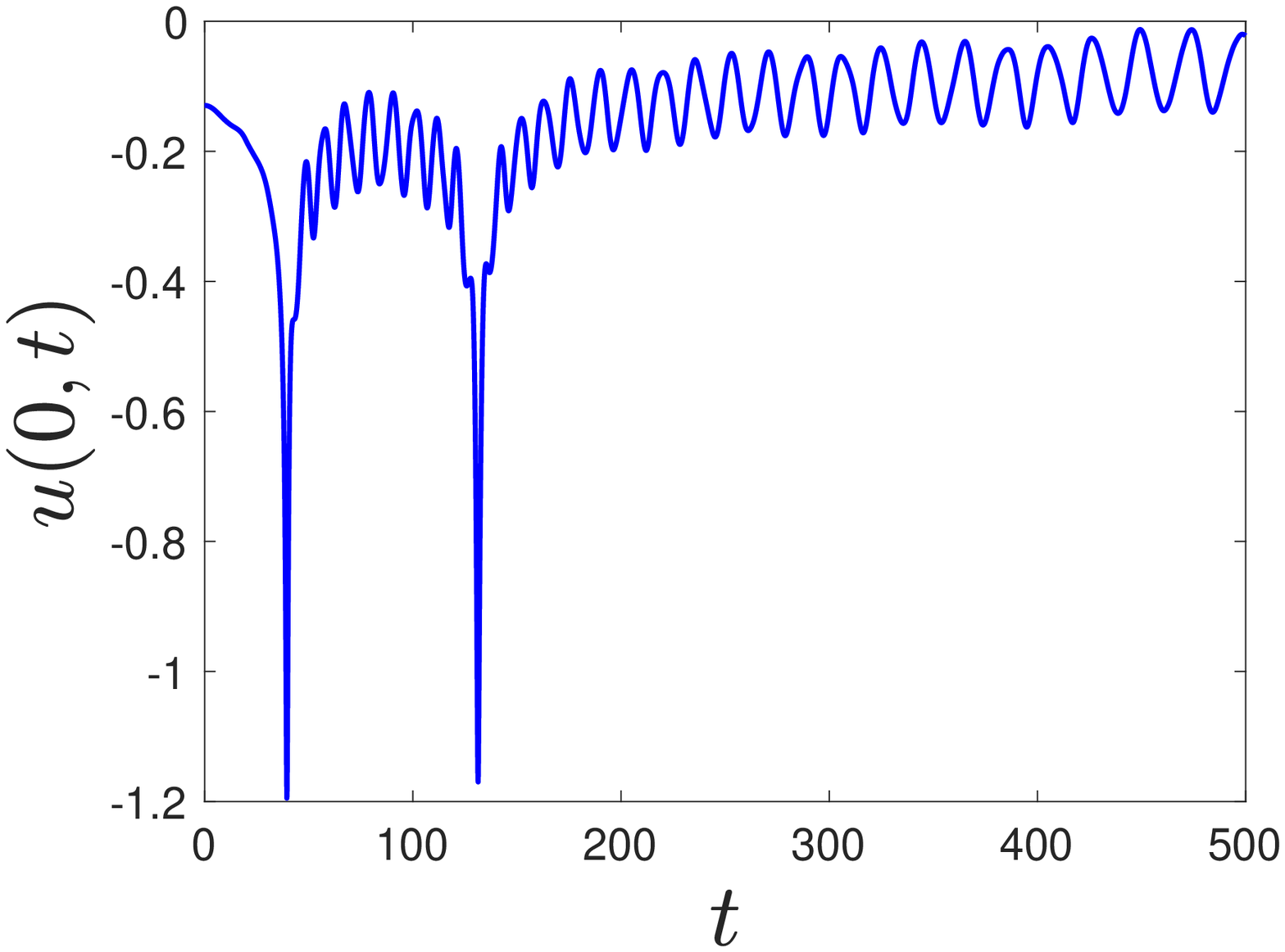}}}
\caption{Space-time contour plot ($u(x,t)$) of the kink-antikink interaction in the $\phi^8$ model for (a) $v_{\rm in}^{}=0.13$ and (c) $v_{\rm in}^{}=0.1357$. Position plot of the field's value at the center point,  $u(x=0,t)$  for (b) $v_{\rm in}^{}=0.13$ corresponds to $M=8$, and (d) $v_{\rm in}^{}=0.1357$ corresponds to $M=9$. Here $M$ counts the number of small oscillations of  $u(x=0,t)$ occurring inside a two-bounce window. Notice that the ``small'' bounces seen in the center-point plots (right panels) are reflected in the alternating dark/light vertical lines in the contour plots (left panels).}
\label{twoBounceViewsPhi8}
\end{figure}

We now dig deeper into the behavior of the two-bounce windows. In Fig.~\ref{twoBounceViewsPhi8}
we look at space-time contour plots and center-point field value $u(x=0,t)$ versus time plots for the $\phi^8$ model. Following Ref.~\cite{Campbell.PhysD.1983}, we define a ``bounce number'' $M$ that counts the number of small oscillations of the field at the center point, i.e., $u(x=0,t)$, that occur inside a two-bounce window. In the contour plots these small oscillations appear as alternating light and dark vertical stripes, and in the center-point plots they appear as alternating relative maxima and minima. For this we count both of the large bounces that define the two-bounce window as well as the small oscillations that occur between the two large bounces. For example, in Fig.~\ref{twoBounceViewsPhi8}(b), we count the larger local minima at about $t=40$ and $t=120$ and the six smaller local minima in between (we do not count the very small ``blips'' near the large local minima) for a value $M=8$. We can also refer to this window as the $n=2$ two-bounce window, since it is represented by the second blue curve in Fig.~\ref{vin_vout-8}. We call $n$ the window number.

\begin{figure}[h]
\centering
\subfigure[]{\includegraphics[width=0.46\textwidth]{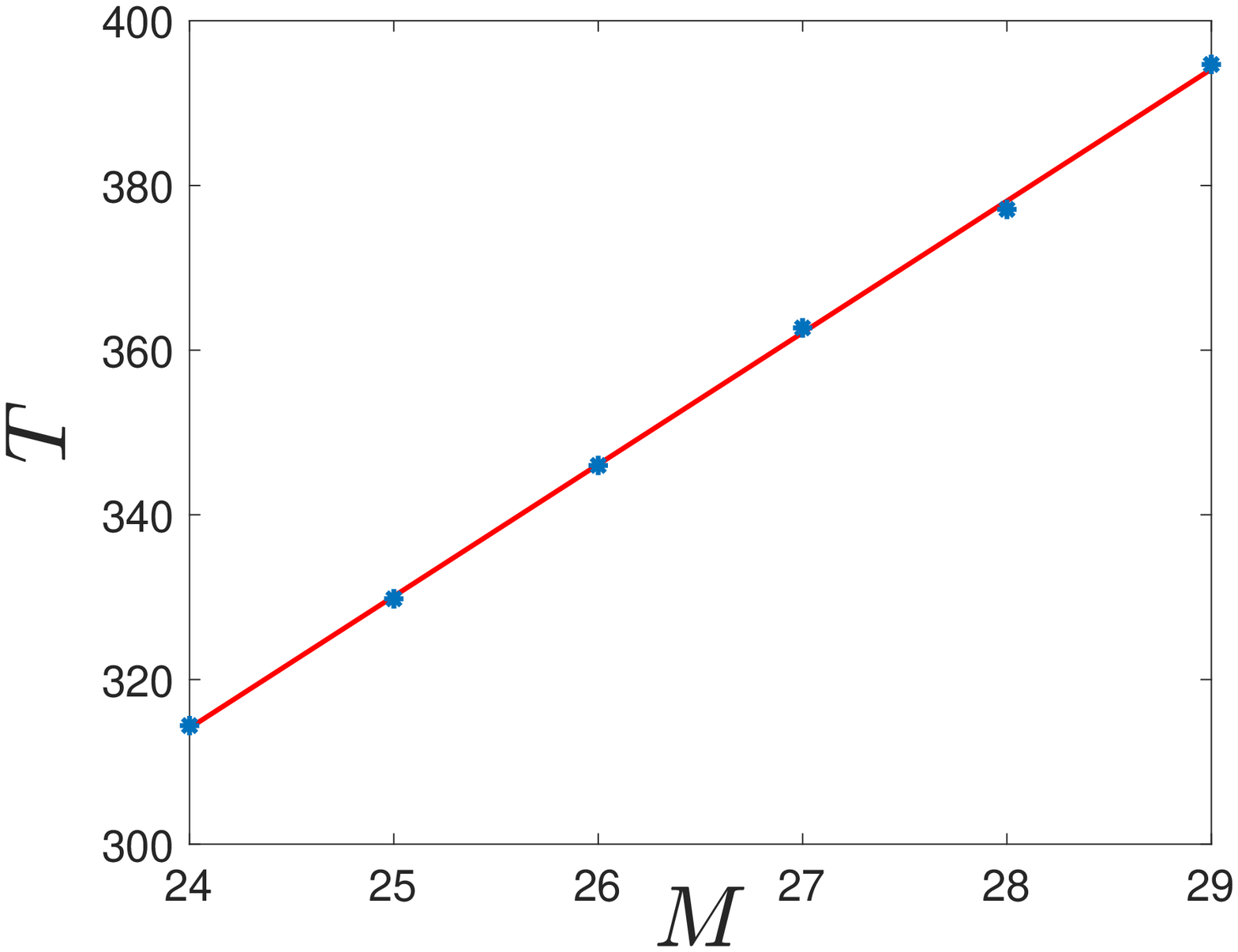}}%
\subfigure[]{\includegraphics[width=0.46\textwidth]{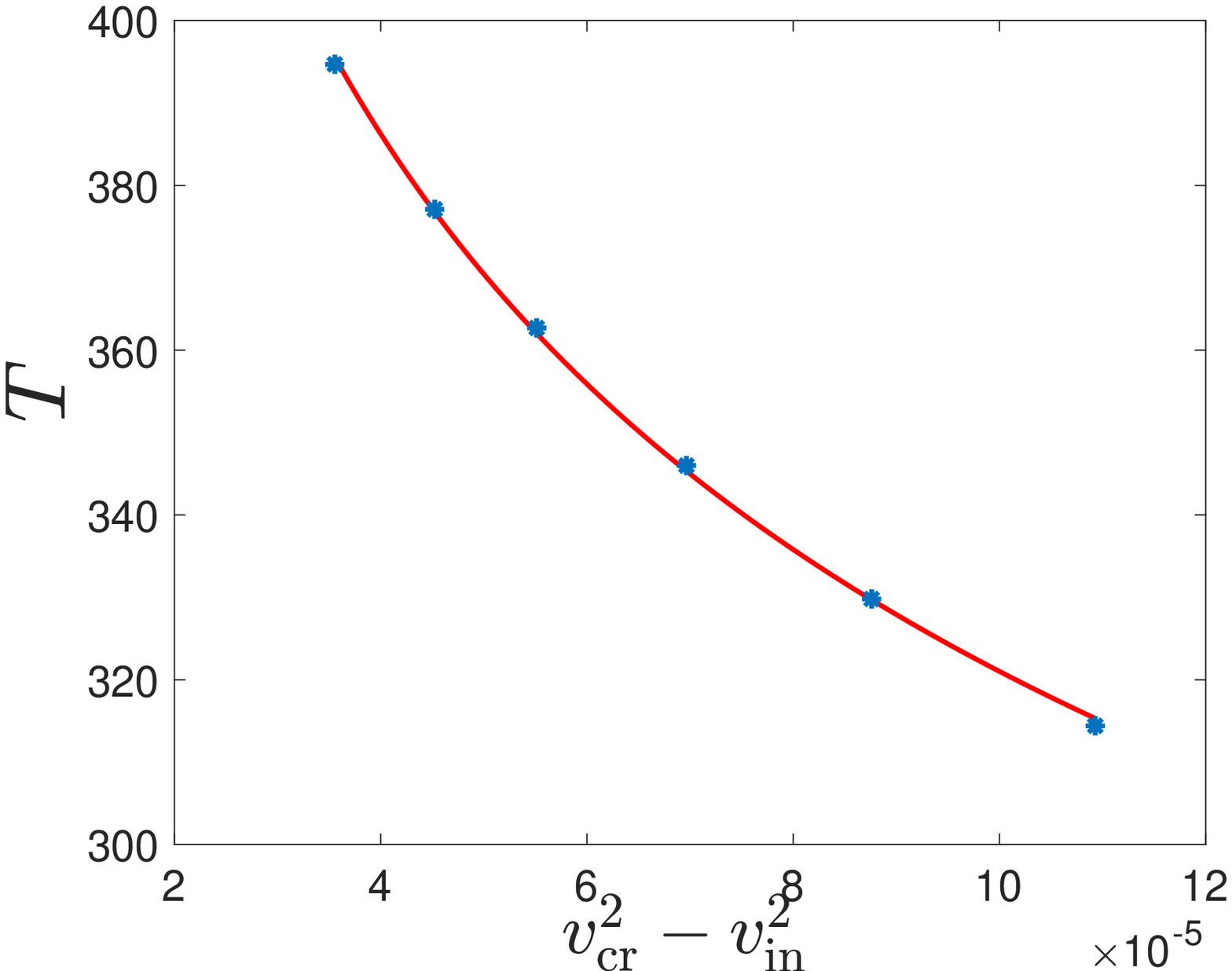}}
\caption{(a) Width $T$ of the two-bounce window versus the bounce number $M$ for the $\phi^8$ model. (b) $T$ versus $v_{\rm cr}^2-v_{\rm in}^2$ curve.}
\label{TMandTVfit}
\end{figure}

Next, we define the width $T$ of a two-bounce window as $T_2-T_1$ where $T_1$ and $T_2$ are the $t$-values where the larger local minima occur in the center-bounce plots (the $t$-width of the bounce interval). For example, in Fig.~\ref{twoBounceViewsPhi8}(b), we have $T_1=40.5$ and $T_2=117.8$. Now we assume $T=aM+b$ and $T=C\left(v_{\rm cr}^2-v_{\rm in}^2\right)^k$ where $v_{\rm in}^{}$ is the midpoint of the interval that results in bounce number $M$. This is inspired by the approach in Ref.~\cite{Campbell.PhysD.1983}, except the latter took $T=C\left(v_{\rm cr}^2-v_{\rm in}^2\right)^{-1/2}$, and they used the window number $n$ in place of the bounce number $M$. Using the windows with $M$ values $24$ through $29$, the fitted equations are $T=16.00M-69.96$ and $T=49.98(v_{\rm cr}^2-v_{\rm in}^2)^{-0.2019}$. These relationships are assumed to apply for windows that are close to $v_{\rm cr}^{}$, and so we only use the last few $M$ values that we calculated. Figure \ref{TMandTVfit} shows the quality of these curve fits. Combining these equations results in
\begin{equation}
v_{\rm in}^{} = \sqrt{v_{\rm cr}^2-\left( \frac{aM+b}{C}\right)^{1/k}} \,.
\label{vinPredict1}
\end{equation}
For our particular parameter values:
\begin{equation}
v_{\rm in}^{}(M)=\sqrt{0.02247-\left(0.3202M-1.400\right)^{-4.952}} \,.
\label{vinPredict2}
\end{equation}

Using Eq.~\eqref{vinPredict2}, we were able to predict the midpoints of some two-bounce intervals that our original search method missed, showing the promise of this approach. Consequently, we now have two-bounce windows for the $\phi^8$ model for $M$ values $7\le M\le 29$.
One major difference between the collision phenomenology in the chosen example $\phi^8$ model and the classical $\phi^4$ field theory appears to be that, for the $\phi^8$ case, the first two-bounce window occurs for $M=7$, whereas for the $\phi^4$ case the first two-bounce window occurs for $M=3$. We find that $n=M-6$ for the $\phi^8$ model (instead of $n=M-2$ for the $\phi^4$ model). For the $\phi^8$ model, following the first two-bounce window at $M=7$, the second two-bounce window occurs at 8 oscillations, the third at 9 oscillations, and so on, keeping the $n=M-6$ pattern. We have shown that this pattern continues up to at least $M=29$ oscillations (i.e., window number $n=23$).

\subsection{$\phi^{10}$ and $\phi^{12}$ cases}

\begin{figure}[t!]
\centering
\subfigure[]{{\includegraphics[width=0.49\textwidth]{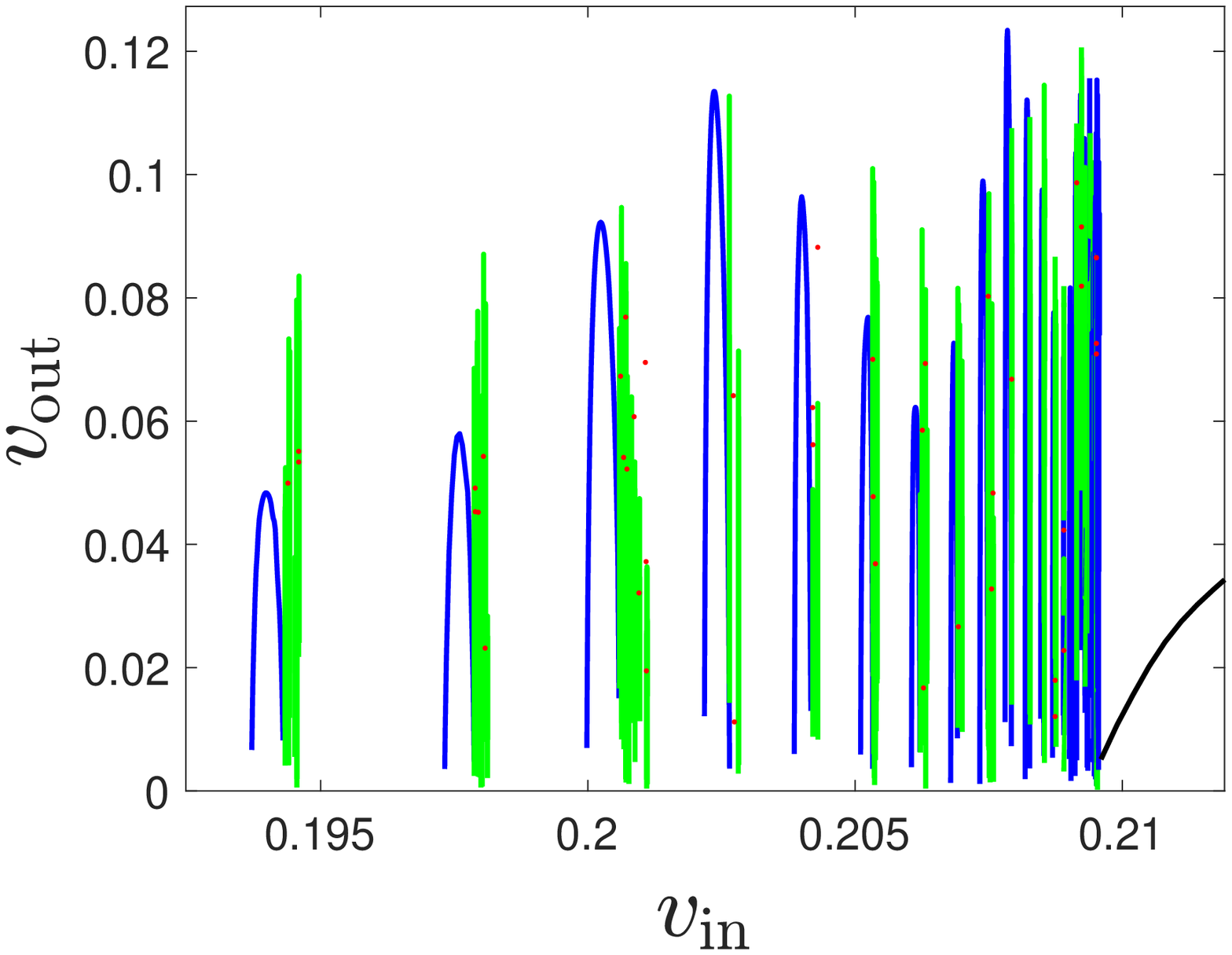}}} \subfigure[]{{\includegraphics[width=0.49\textwidth]{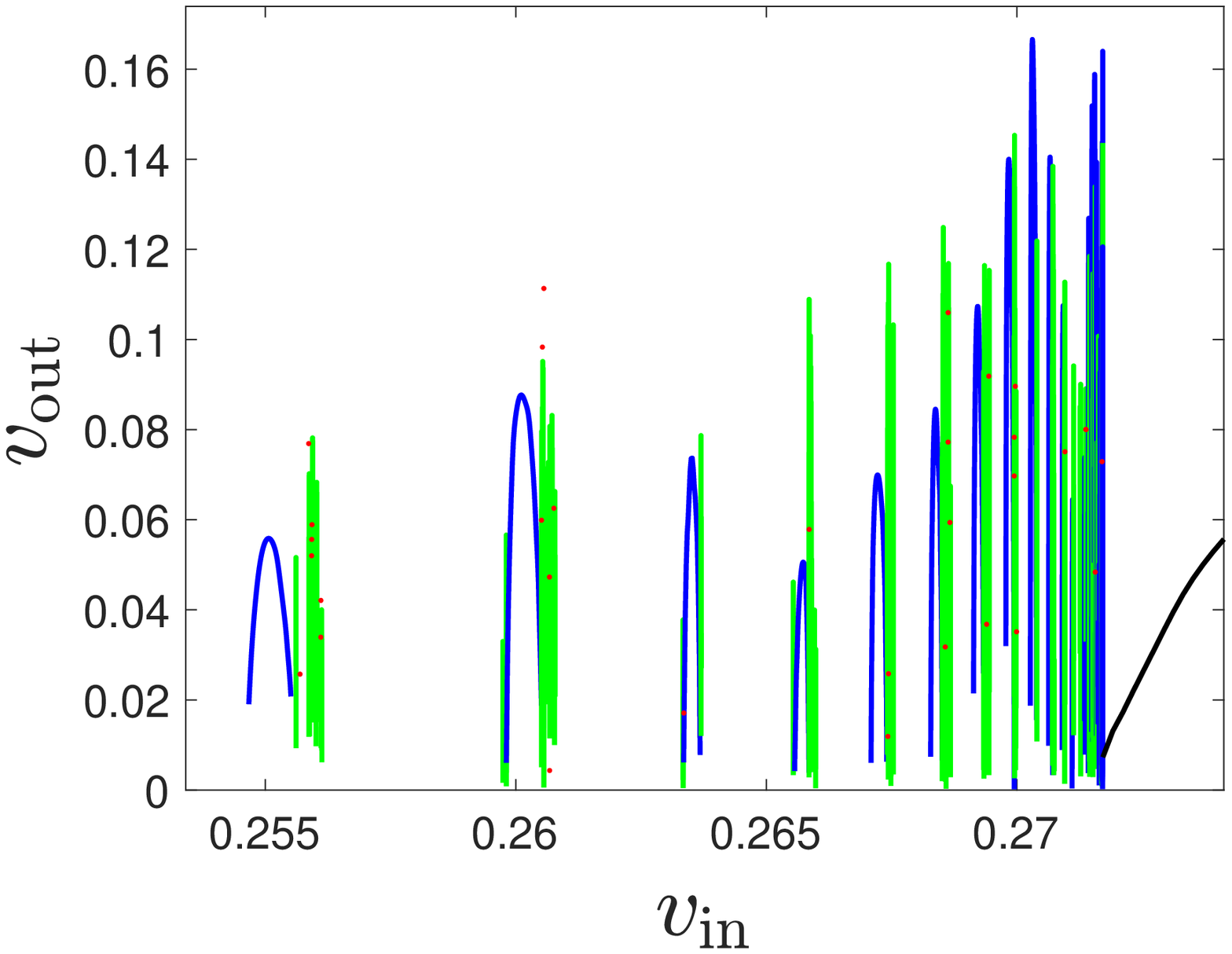}}}
\caption{The escape velocity $v_{\rm out}^{}$ as a function of the initial velocity $v_{\rm in}^{}$ for (a) the $\protect\phi^{10}$ model and (b) the $\protect\phi^{12}$ model. Black represents a one-bounce solution, blue represents a two-bounce solution, green represents a three-bounce solution, and red (dots) represents a four-bounce solution.}
\label{vin_vout-1012}
\end{figure}

The $v_{\rm in}^{}$ -- $v_{\rm out}^{}$ graphs for the $\phi^{10}$ and $\phi^{12}$ models show the same fractal-like structure and alternately increasing and decreasing maximum values as for the $\phi^8$ model, see Fig.~\ref{vin_vout-1012}. Similar to the $\phi^8$ case we were able to find equations for the $\phi^{10}$ and $\phi^{12}$ cases that relate the bounce number $M$ to the midpoint of the corresponding window (in terms of $v_{\rm in}^{}$), using the windows with values $M=24$ through $M=29$. Again, we assume $T=aM+b$ and $T=C\left(v_{\rm cr}^2-v_{\rm in}^2\right)^k$, fit the parameters to the data, and obtain equations in the form of Eq.~\eqref{vinPredict1}. 

The graphs comparing the data to the fitted equations are similar to Fig.~\ref{TMandTVfit}. For the $\phi^{10}$ model, we obtain the fits $T=24.06M-142.5$ and $T=58.78\left(v_{\rm cr}^2-v_{\rm in}^2\right)^{-0.2407}$, and for the $\phi^{12}$ model we obtain the fits $T=24.87M-132.8$ and $T=85.80\left(v_{\rm cr}^2-v_{\rm in}^2\right)^{-0.1810}$. 
These fits are highly accurate descriptions of the underlying data (the coefficient of determination is $r^2\approx 0.999$). The equation that predicts the location of a two-bounce window with a given $M$ value for the $\phi^{10}$ model (and that is analogous to Eq.~\eqref{vinPredict2} for the $\phi^8$ model) is
\begin{equation}
v_{\rm in}^{}(M) = \sqrt{0.04393-\left(0.4093M-2.425\right)^{-4.155}} \,. \label{vinPredict3}
\end{equation}
Meanwhile, for the $\phi^{12}$ model, the fit is
\begin{equation}
v_{\rm in}^{}(M) = \sqrt{0.07383-\left(0.2899M-1.5478\right)^{-5.526}} \,.  \label{vinPredict4}
\end{equation} 
The non-monotonicity of the relevant exponents as the order
of the field theory gets higher is a topic meriting further investigation.

Finally, for the $\phi^{10}$ model, the first window ($n=1$) has bounce number $M=10$ and the pattern $n=M-9$ continues for all cases that we considered. For the $\phi^{12}$ model, the first window has bounce number $M=8$; the pattern $n=M-7$ continues up through the $M=19$ window. Surprisingly, there is no $M=20$ window; this is the only case encountered so far of a ``missing'' window. Window numbers $M=21$ through $M=36$ have been shown to exist (numerically). We will not pursue this further here, but it appears that the missing window occurs at a point in the window pattern where the local maximum of the $v_{\rm in}^{}$ -- $v_{\rm out}^{}$ graph would be very small (if it existed, based on nearby windows). This could also help to explain why there are ``missing'' windows at the beginning of the sequence of two-bounce windows, where the local maxima are small (the first two-bounce window could theoretically be as small as $M=2$). 

It is important to highlight that while these relations are phenomenological in nature (they were introduced as such even in the pioneering work of Ref.~\cite{Campbell.PhysD.1983}), nevertheless, there as well as here, they are found to be in {\it extremely good} agreement with the numerical results. This, in turn, is a finding that suggests further mathematical investigation along this vein. However, equally importantly, for our purposes, these relations do hold predictive value as while they were only identified for a range of $M$ values, they were confirmed to be very accurate for a far wider range of $M$ values. In fact, in some cases, the delicate nature of the corresponding computations had not revealed the associated window during a first parametric search, yet this prediction enabled a finer search that eventually identified the relevant resonance.

\subsection{Once-minimized versus twice-minimized initial conditions}

\begin{figure}[t!]
\centering
\subfigure[]{\includegraphics[width=0.46\textwidth]{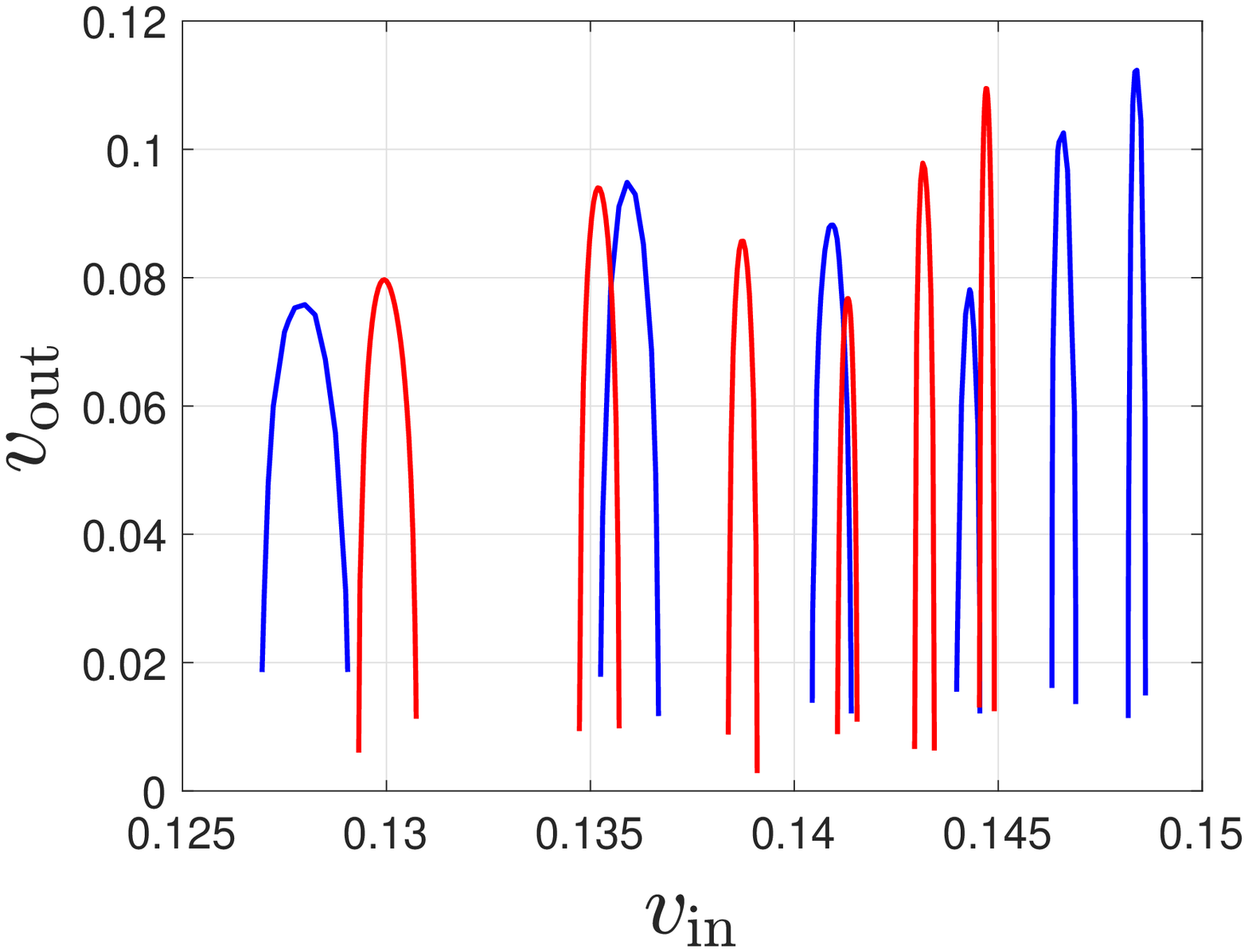}}\subfigure[]{\includegraphics[width=0.46\textwidth]{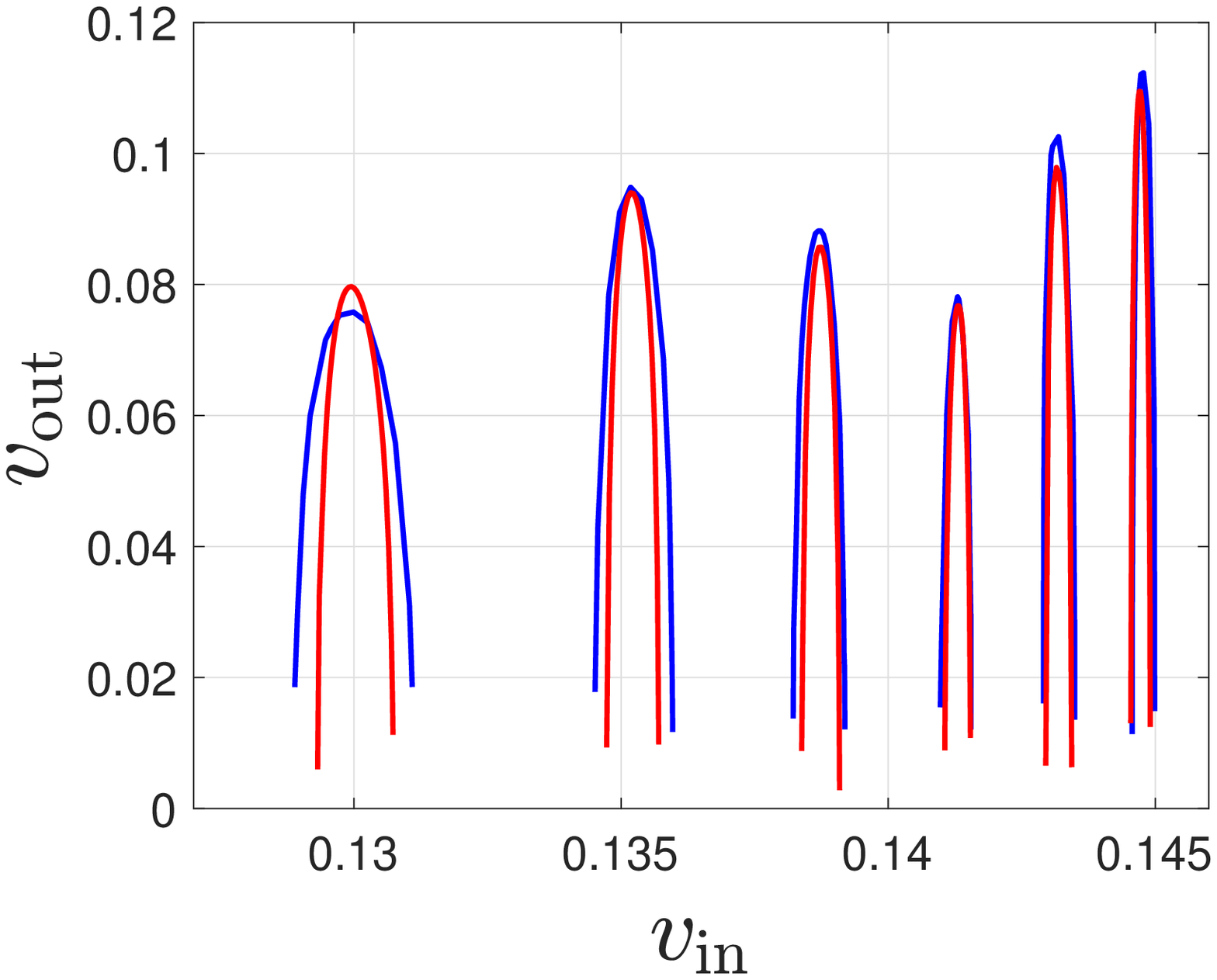}}
\caption{The $v_{\rm in}^{}$ -- $v_{\rm out}^{}$ relationship for the first six two-bounce windows of the $\phi^8$ model. (a) Once-minimized initial conditions in red, twice-minimized in blue. (b) Same as (a) but the blue curves have been shifted to the left and then scaled.}
\label{shifted}
\end{figure}

We now come back to the issue of how the $v_{\rm in}^{}$ -- $v_{\rm out}^{}$ graphs for once-minimized versus twice-minimized initial conditions (as explained in Section \ref{sec:Initial_Conditions}) for the $\phi^8$ model are related. In Fig.~\ref{shifted}(a),
we show the $v_{\rm in}^{}$ -- $v_{\rm out}^{}$ curves for both cases; once-minimized in red and twice-minimized in blue. The critical value $v_{\rm cr}^{}$, where the one-bounce interval begins, is slightly different for the once- versus twice-minimized initial conditions. We first shifted the twice-minimized windows by the amount equal to the difference in the two critical values so that the windows were accumulating at the same value for both cases. We then scaled the $v_{\rm in}^{}$ values of each blue window by a factor $H$ that is dependent on the bounce number $M$, given by $H=11.51\exp(-0.6877M)+1$ (obtained using judicious curve fitting). The point here is not so much the exact amount of scaling and shifting, but rather that the $v_{\rm in}^{}$ -- $v_{\rm out}^{}$ curves using the improved initial conditions (twice-minimized) can be mapped to the corresponding curves for the once-minimized initial conditions, thereby justifying the use of the somewhat ``imperfect'' initial conditions to generate the large amount of data required to construct these curves. Figure \ref{shifted}(b) shows the result of shifting and scaling. The shifting and scaling factors are dependent on the potential of the model, that is, these particular factors apply only to the $\phi^8$ model. Similar transformations can be obtained for the $\phi^{10}$ and $\phi^{12}$ models.

\subsection{Three-bounce intervals}

\begin{figure}[t!]
\centering
\subfigure[]{{\includegraphics[width=0.49\textwidth]{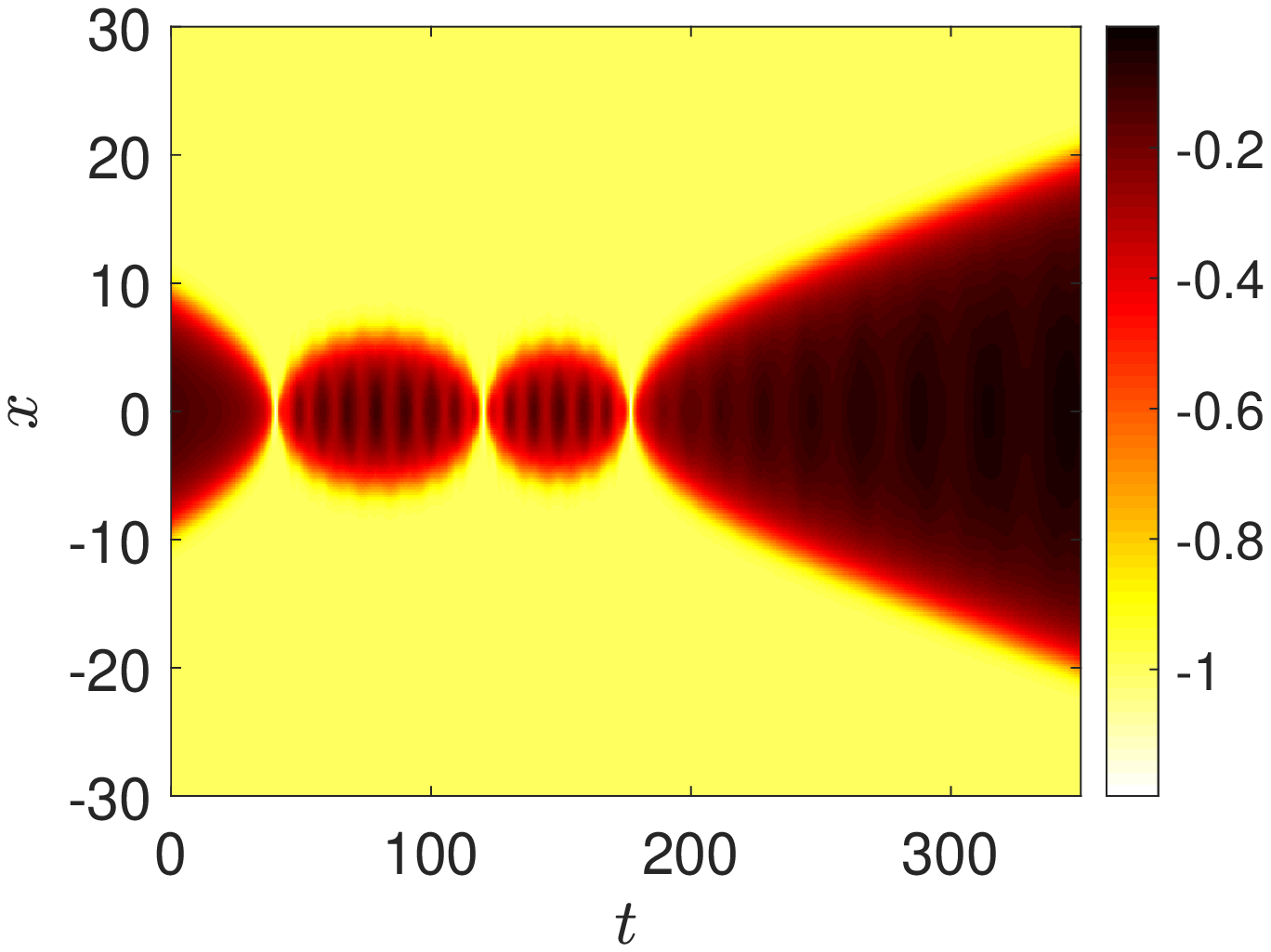}}} \subfigure[]{{\includegraphics[width=0.49\textwidth]{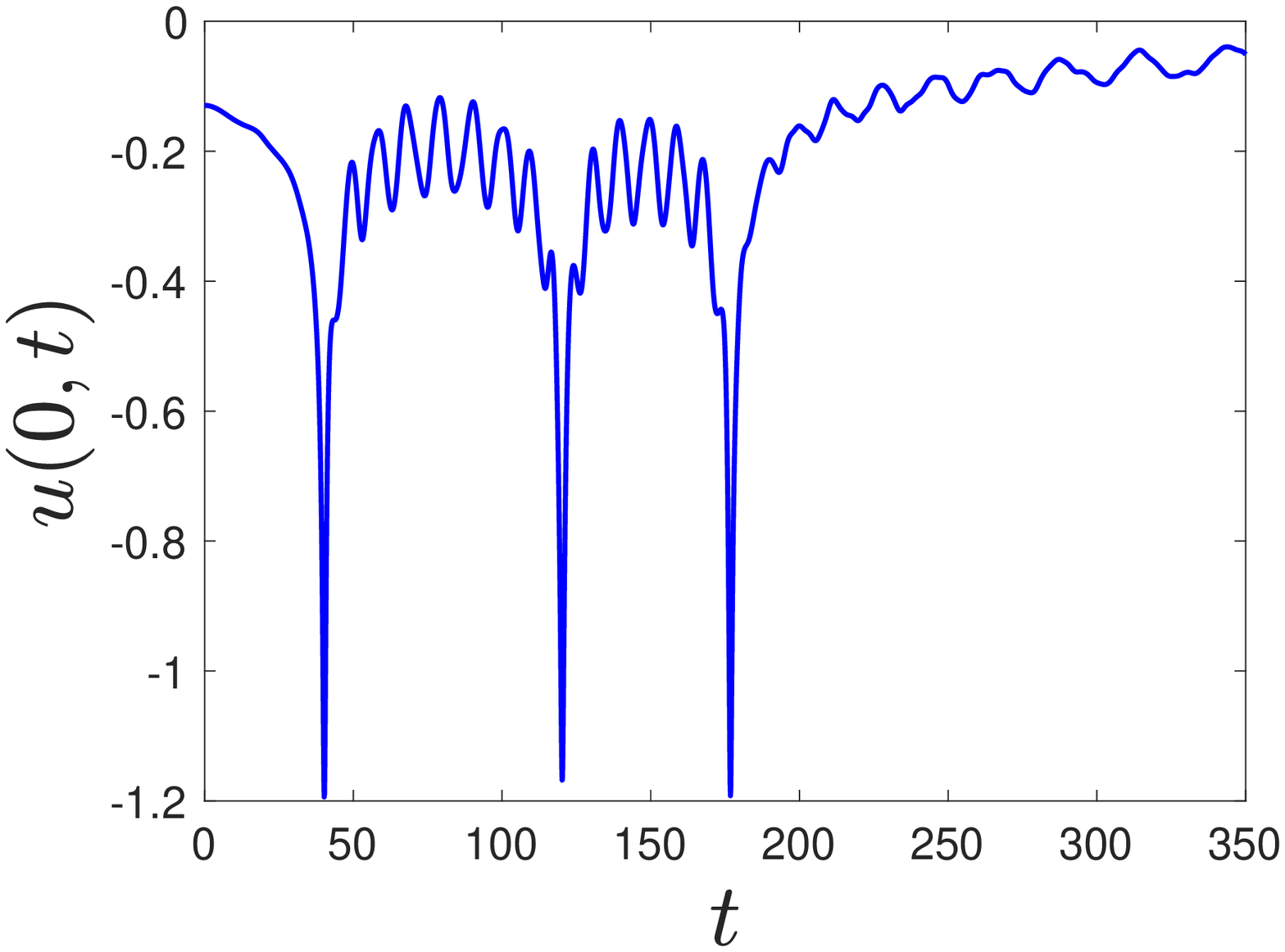}}}
\subfigure[]{{\includegraphics[width=0.49\textwidth]{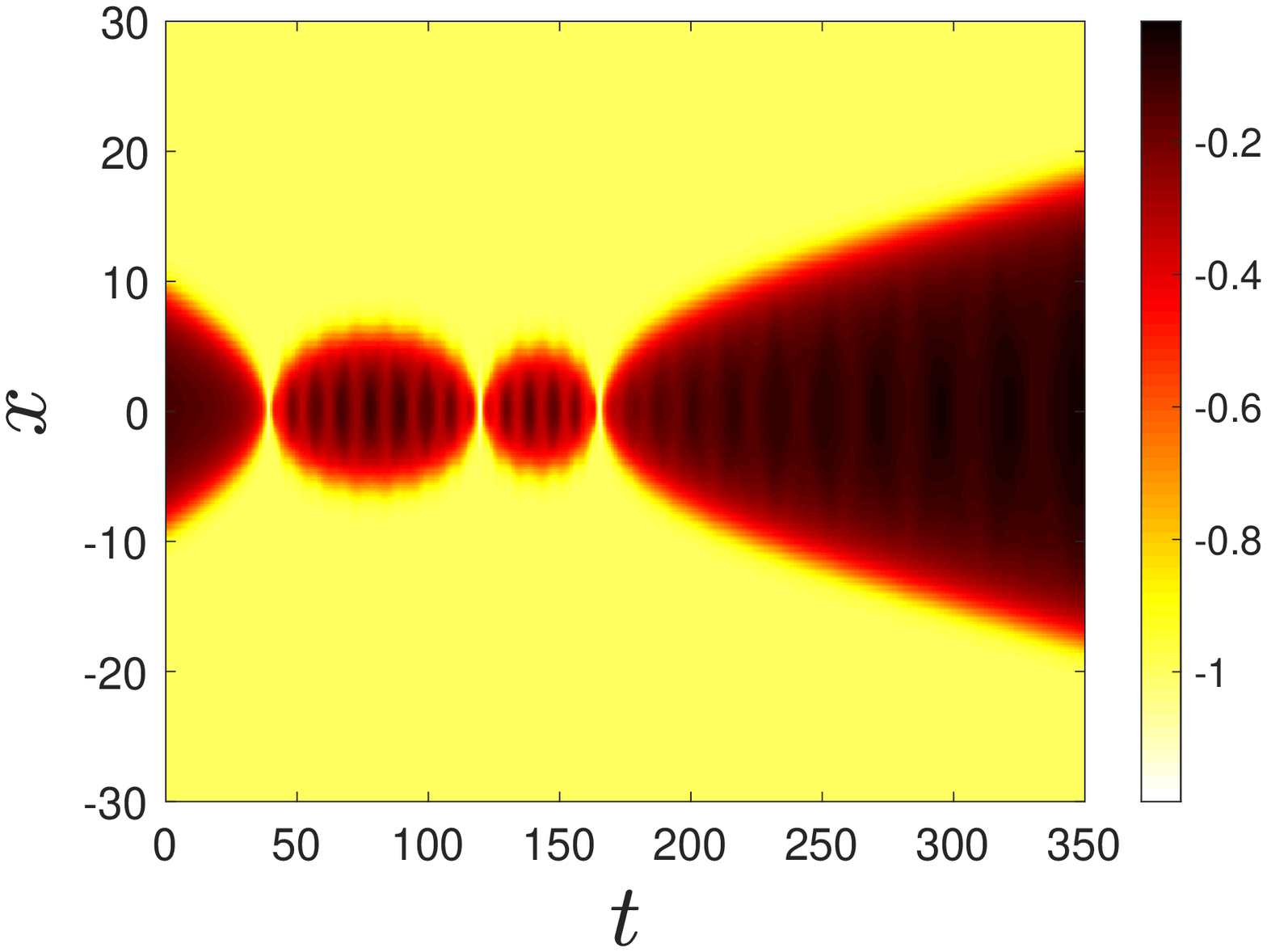}}} \subfigure[]{{\includegraphics[width=0.49\textwidth]{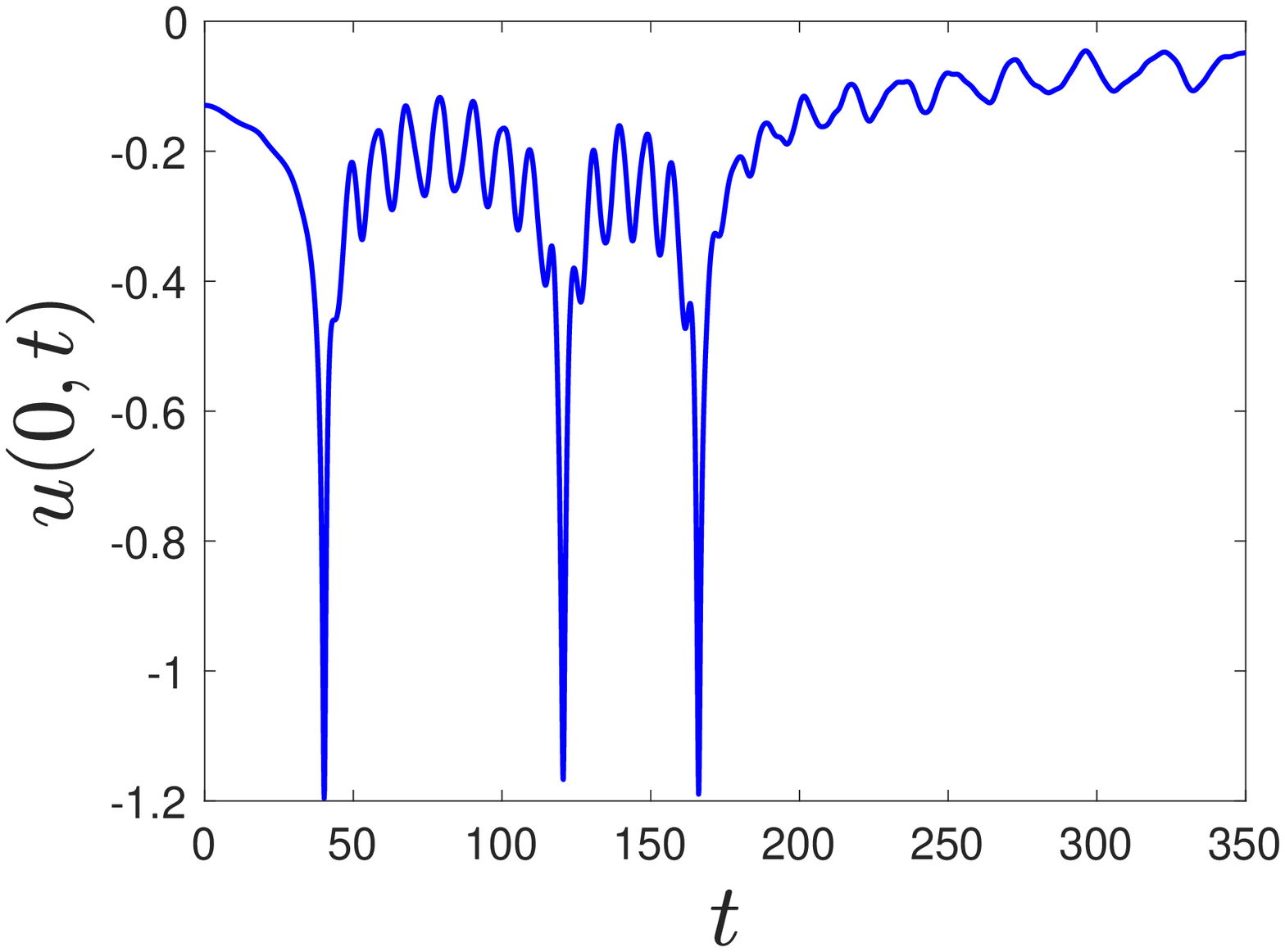}}}
\caption{Space-time contour plot of $u(x,t)$ for the $\phi^8$ model with three-bounce windows: (a) $v_{\rm in}^{}=0.131275$, (c) $v_{\rm in}^{}=0.13143$. Time dependence of the field's value at the center point, $u(x=0,t)$: (b) $v_{\rm in}^{}=0.131275$ corresponding to $M=6$, and (d) $v_{\rm in}^{}=0.13143$ corresponding to $M=5$.  The ``small'' bounces seen in the center-point plots (right panels) are reflected in the alternating dark/light vertical lines in the contour plots (left panels)}
\label{threeBounce}
\end{figure}

The structure of three-bounce windows shows a number of similarities to the structure of two-bounce windows. In Fig.~\ref{threeBounce},
we show contour plots and center-point plots for two three-bounce windows for the $\phi^8$ model. An $M$ value can be defined for the three-bounce windows, as for the two-bounce windows, by counting the number of small oscillations of the field between the  second and the third bounces. In Fig.~\ref{threeBounce}(a,b), we count $M=6$ for the first case, while we count $M=5$ for the case in panels (c,d). Using these $M$ values, we can perform curve fitting to obtain functional relationships, as for the two-bounce windows, using the edges of the two-bounce window as $v_{\rm cr}^{}$. 

The resulting equations can be used to predict new three-bounce windows with some success. One major difference, compared to two-bounce windows, is that three-bounce windows that accumulate at the left of a two-bounce window appear to be unique in the same sense as two-bounce windows; no two windows have the same bounce number $M$. However, the three-bounce windows that accumulate at the right of a two-bounce window are not unique; there can be separate three-bounce windows with the same $M$ value. 
This issue merits further investigation. 

\section{Corresponding Initial Conditions at $t=-\infty$}
\label{sec:Correspondence}

To investigate kink-antikink interactions, we need to choose the same initial half-separation for all of the runs. Because of the polynomial tails (which face each other) we can never achieve a large enough separation to effectively simulate an initial infinite separation (the simulation run times would be too long). We settled on using an initial half-separation of $x_0^{}=10$. Every initial kink velocity $v_{\rm in}^{}$ at $x_0^{}=\infty $ will correspond to some velocity at $x_0^{}=10$ and vice versa, thus there is a one-to-one correspondence between the velocities of a kink at any two $x_0^{}$ values (including infinity). Then, when we determine quantities like critical $v_{\rm cr}^{}$ and the various bounce windows (two-bounce, three-bounce, and so on) in terms of $v_{\rm in}^{}$ at $x_0^{}=10$, we can, if we want, calculate the velocities these would correspond to at a different separation, including, possibly, an infinite
one.

In Ref.~\cite{our_PRL}, accelerations (starting from rest) between kink and antikink were numerically calculated for various half-separation distances $x_{0}$. For the $\phi^8$ model, as long as $x_{0}\geq 50$, these numerically calculated accelerations agree with those calculated from theory (``Manton's method'' \cite{Manton.NPB.1979,Kevrekidis.PRE.2004}) to within $0.4\%$. In fact, as the separation distance becomes larger, the error decreases (presumably, to zero). Thus for large separations, we can use the ODE $x^{\prime\prime}=-11.08/x^4$ \cite[Table II]{our_PRL} corresponding to the $\phi^8$ model. That is, taking a given initial velocity at a given $x_0^{}$, we can run it backwards $(t\rightarrow -\infty )$ to obtain the $v_{\rm in}^{}$ that corresponds to a different (or even possibly infinite) initial separation. The equation $x^{\prime\prime}=-11.08/x^4$  integrates to $x^{\prime 2}=\frac{22.16}{3}x^{-3}+C$; then, letting $t\rightarrow -\infty$ (and hence $x\rightarrow \infty$), we find $x^{\prime }(-\infty)=\sqrt{C}=\sqrt{x^{\prime }(0)^{2}-\frac{22.16}{3x(0)^3}}$. Letting $v_{\rm in}^{}(x_0^{})$ represent the $v_{\rm in}^{}$ value at an initial half-separation of $x_0^{}$, and $v_{\rm in}^{}(\infty)$ be the corresponding $v_{\rm in}^{}$ value at an infinite separation, we have
\begin{equation}
v_{\rm in}^{}(\infty ) = \sqrt{v_{\rm in}^2(x_0^{})-\frac{22.16}{3x_0^3}}.
\label{v_infinity}
\end{equation}
For the $\phi^{10}$ and $\phi^{12}$ models, the same reasoning applies (and so similar formulas can easily be derived), however, the acceleration formulas of Ref.~\cite{our_PRL} for these cases will not be accurate to the degree of the $\phi^8$ case until the half-separation becomes much larger than $x_0^{}=50$.

As shown in Ref.~\cite{our_PRL}, for separations smaller than those for which the acceleration formulas become accurate, we can still use the methods of that paper to generate a very accurate ODE whose solution tracks the centers of the kink and antikink for virtually any separation distance. We do this by creating an interpolating function $a(x)$ that gives the correct acceleration for different separations
using data points for acceleration versus half-separation $x$ (calculated numerically as in Ref.~\cite{our_PRL} for $x$ from $50$ to $300$). Then, the ODE $x^{\prime\prime}(t)=a(x(t))$ tracks the position of the kinks of the PDE quite well (even for small $x(t)$).

\begin{figure}[t!]
\centering
\subfigure[]{\includegraphics[width=0.46\textwidth]{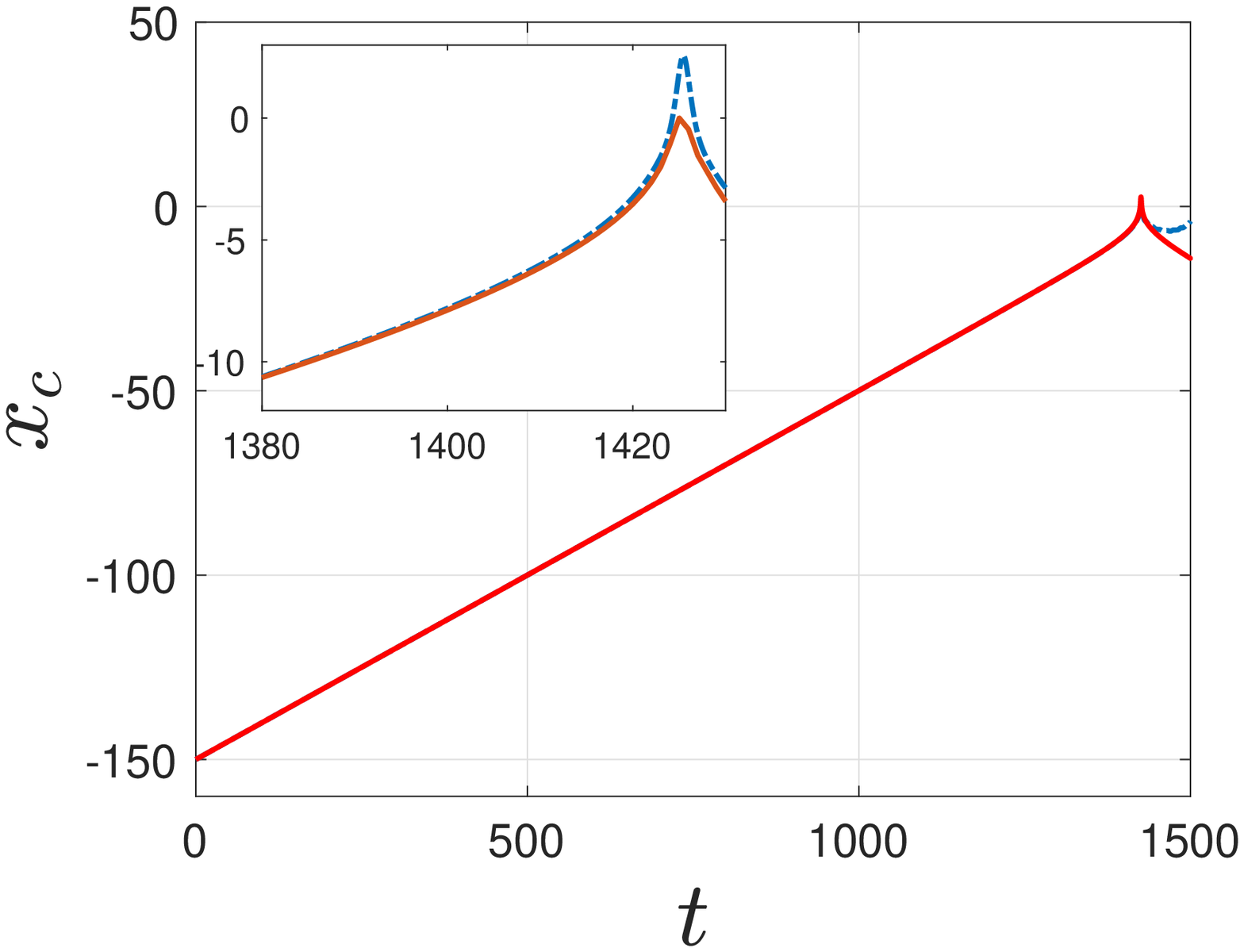}}\subfigure[]{\includegraphics[width=0.46\textwidth]{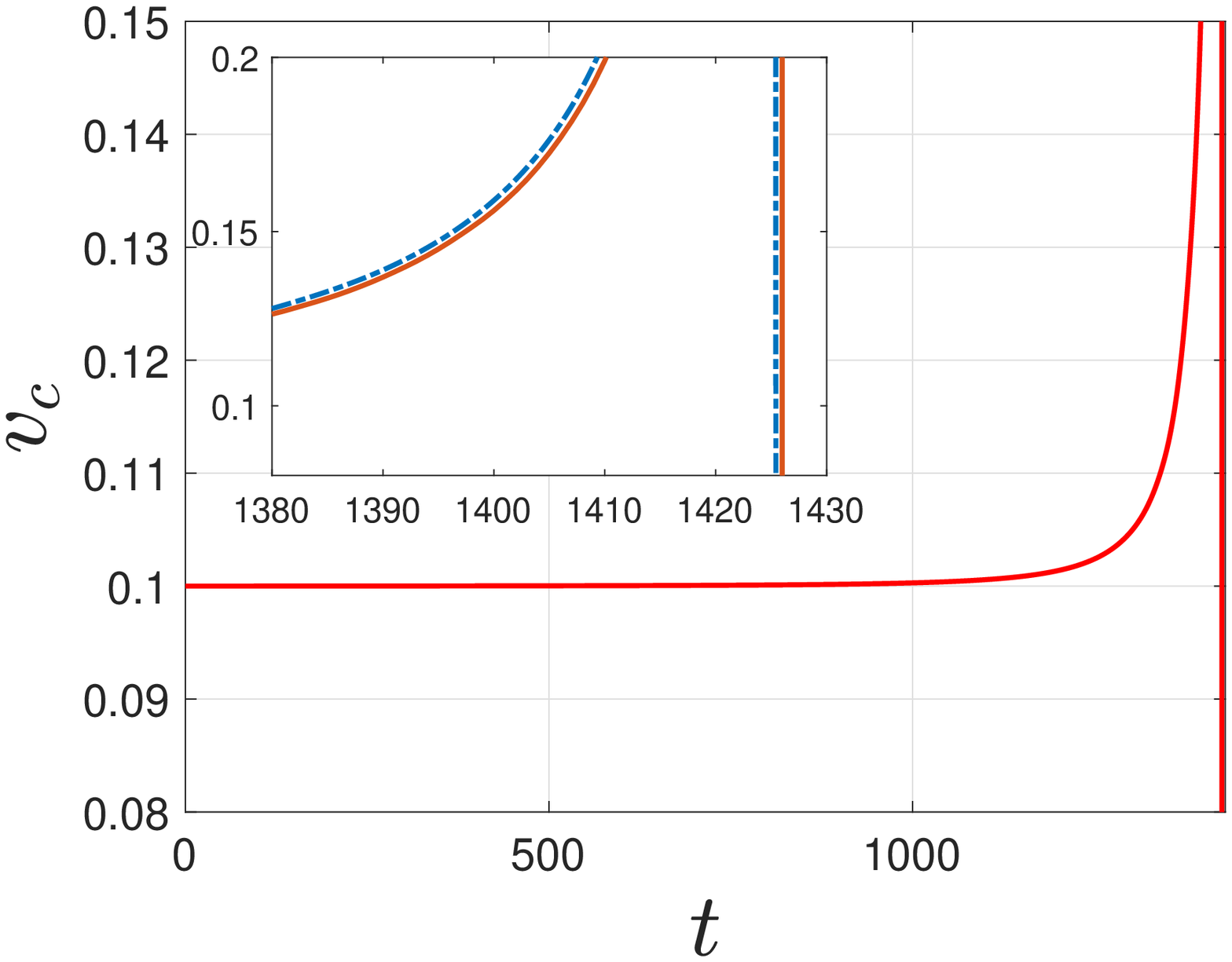}}
\caption{(a) Plot of the kink's center point $x_{\rm c}(t)$, and (b) its velocity $v_{\rm c}(t)$ for the $\phi^8$ model. The PDE solution is shown as the blue dashed curve, while the ODE solution is shown as the solid red curve. Insets show magnification of the plots near the ``peaks''.}
\label{ODEvsPDEinset}
\end{figure}

In Fig.~\ref{ODEvsPDEinset},
we show the position and velocity of the $\phi^8$ kink, with initial half-separation $x_0^{}=150$ and initial velocity $v_{\rm in}^{}=0.1$, as calculated by the PDE, versus position and velocity as calculated by the ODE $x^{\prime\prime}(t)=a(x(t))$, showing that the PDE and ODE give nearly identical results over a large scale. The insets show zooms indicating where the two methods diverge somewhat, though still very little, for separations less than $10$.
Of course, it is relevant to remind the reader that these ODE results cannot match the PDE results after the point of interaction between kink and antikink. 
Given the absence of internal mode considerations in this ODE, it is not possible
for the latter to reflect, e.g., the multi-bounce window behavior of the PDE.

\section{Conclusions, Challenges, and Future Work}
\label{sec:Conclusion}

In this paper, we have investigated kink-antikink collisions in the $\phi^8$, $\phi^{10}$, and $\phi^{12}$ models. To accomplish this, first of all, we asked ourselves how to correctly formulate the initial conditions in a situation when the kinks face each other with power-law tails. The main problem is that the power-law kinks' tails overlap significantly at any acceptable finite initial separation. As a consequence, it is far less
straightforward to select an
initial distance that can be considered as ``infinitely large'', i.e., large enough 
so that the kink and antikink could be considered non-interacting.

In Ref.~\cite{our_PRD}, a ``minimization'' procedure was developed to allow us to obtain the {\it static} ``kink+antikink'' configuration at any finite initial separation distance, and to ensure that this configuration is a solution of the equation of motion. In this paper, we advanced the approach further and showed how to build initial conditions in the form of a kink and an antikink moving towards each other with a given \emph{nonzero} initial velocity. We started with the split-domain ansatz \cite[Sec.~III.D]{our_PRD} and minimized it for the case of a moving kink and antikink configuration. Thus, we obtained the $x$-dependence of the field at $t=0$, i.e., $u(x,0)$. To numerically solve the equation of motion, however, we also need the $x$-dependence of the time derivative of the field at $t=0$, i.e., $u_t(x,0)$. We proposed two different methods for obtaining appropriate initial conditions for the numerical simulation of the kink-antikink collisions. The two proposed approaches differ in the method of obtaining $u_t(x,0)$: {\it once-minimized} and {\it twice-minimized}.

Using the once-minimized initial conditions, we investigated the collisions of a kink and an antikink in the case of the $\phi^8$, $\phi^{10}$, and $\phi^{12}$ field-theroretic models in which the kinks exhibit power-law tails. We constructed a typical picture of the interactions: there is a critical initial velocity $v_{\rm cr}^{}$ such that for $v_{\rm in}^{}>v_{\rm cr}^{}$ the kinks escape to infinity after a single impact. At $v_{\rm in}^{}<v_{\rm cr}^{}$ the kinks' capture leads to the formation of a bound state. Besides that, two-bounce, three-bounce, and so on escape windows were found in the range $v_{\rm in}^{}<v_{\rm cr}^{}$. For the $\phi^8$ model, the critical velocity is $v_{\rm cr}^{}\approx 0.15$, while for the $\phi^{10}$ and $\phi^{12}$ models the respective values are $v_{\rm cr}^{}\approx 0.2096$ and $v_{\rm cr}^{}\approx 0.2717$.  Using a fit similar to the one in Ref.~\cite{Campbell.PhysD.1983}, we obtained formulas that relate the incoming velocity associated with a two-bounce window ($v_{\rm in}^{}$ in the middle of the window) with $v_{\rm cr}^{}$ and the bounce number $M$, as well as with the time between adjacent collisions. Moreover, for the $\phi^8$ model, a similar analysis can be performed for three-bounce windows (although only for the ones arising to the left edge of their corresponding (adjacent) two-bounce windows). Importantly, the definite power law nature of the associated results (and associated exponents), as well as the wide range (e.g., over $M$) of their applicability suggests that it would be particularly interesting to seek a coherent physical explanation behind such phenomenological relations.

Next, we compared the results of the study of the two-bounce windows in the $\phi^8$ model obtained using the once-minimized and the twice-minimized initial conditions. Despite the fact that these two methods lead to slightly different patterns of the escape windows, we showed that using simple scaling, the results of the two methods can be mapped to each other. This is an important observation, since it means that for calculations one can use a much faster (but less accurate) method based on the once-minimized initial conditions.

Finally, in the $\phi^8$ model, using the formulas for the kink's accelerations obtained in prior work \cite{our_PRL}, a one-to-one correspondence was established between the initial velocities at any finite kink-antikink separation and the initial velocity formally corresponding to an infinitely large initial separation.

As indicated in the Introduction, 
when such multi-bounce collision phenomena arise, it is appealing to
attempt to connect them to collective coordinate
descriptions. Indeed, a long albeit still inconclusive
attempt in that direction is ongoing in the
case of the regular $\phi^4$ model. In the present case,
it was argued that such vibrational modes of the kink could
only exist in the form of embedded eigenvalues within
the continuous spectrum of the problem. 
 Whether such embedded eigenvalues may exist within the continuous spectrum is a mathematically interesting question, meriting further investigation \cite{Gani.JPCS.2020.no-go}. From a physical perspective, a natural follow-up query is: where is a part of the kinetic energy of kinks temporarily accumulated during the time interval between the first and the second impacts within a two-bounce window? The answer may lie in the mechanism proposed in Ref.~\cite{Dorey.PRL.2011} for kinks of the $\phi^6$ model. In Ref.~\cite{Belendryasova.CNSNS.2019}, a similar approach was applied to the kink-antikink system of the $\phi^8$ model. It was shown that in the excitation spectrum of the ``kink+antikink'' system as a whole there are several discrete eigenmodes.
One of these  can be considered to be excited in the kink-antikink collision, this level may accumulate energy, which in turn may lead to the appearance of escape windows in the kink-antikink collisions at $v_{\rm in}^{}<v_{\rm cr}^{}$. However, we caution the reader that such an interpretation involves linearization around a {\it nonstationary} state, so the mathematical foundation of such an approach is presently far from rigorous.

The above considerations also suggest that  the search for the suitable resonance frequency, i.e., the frequency of the vibrational mode into which the kinetic energy of kinks is transferred, remains an open problem to be pursued in future work. Furthermore, knowing the frequency of the vibrational mode could make it possible to investigate the origin of this mode. Once such a (potentially embedded spectrum) mode is identified, this may, in turn, pave the way for developing a reduced description of such algebraically-decaying kinks, on the basis of a suitable choice of collective coordinates \cite{Sugiyama.PTP.1979}. Understanding the degree to which these reductions via collective coordinate approaches may capture the effective phenomenology of the infinite dimensional system, is an interesting research program in its own right and will be addressed carefully in future work.

\section*{Acknowledgments}

The work of the MEPhI group was supported by the MEPhI Academic Excellence Project. V.A.G.\ also acknowledges the support of the Russian Foundation for Basic Research (RFBR) under Grant No.\ 19-02-00971. A.S.\ was supported by the U.S.\ Department of Energy. P.G.K.\ acknowledges support from the US National Science Foundation under Grants No.\ PHY-1602994 and DMS-1809074. He also acknowledges support from the Leverhulme Trust via a Visiting Fellowship and the Mathematical Institute of the University of Oxford for its hospitality during part of this work.

\end{document}